\documentclass[epj]{svjour}
\usepackage[english]{babel}									

\usepackage[utf8]{inputenc}										

\usepackage[T1]{fontenc}									

\usepackage[normalem]{ulem}

\usepackage{amsmath,amsfonts,amssymb,cancel,siunitx,calculator,calc,mathtools,empheq,latexsym}
\usepackage{lineno}

\usepackage{subfig,epsfig,tikz,float}		     
\usepackage{booktabs,multicol,multirow,tabularx,array}   \usepackage{hyperref}       
\usepackage[numbers]{natbib}
\begin{document}

\title{More uses for Thermal Models}
\author{Natasha Sharma\inst{1}, Lokesh Kumar\inst{2} \and Sourendu Gupta\inst{3}
}
\institute{Department of Physics \& Astrophysics, University of Delhi, Delhi 110007, India, \and Department of Physics, Panjab University, Chandigarh 160014, India, \and International Center for Theoretical Sciences, Tata Institute of Fundamental Research, Survey 151 Shivakote, Hesaraghatta Hobli, Bengaluru North 560089, India.}
\date{Received: date / Revised version: date}
\abstract{
 We explore combinations of particle and anti-particle yields which can be used to test thermal models in a parameter free way. We also explore combinations which can be used to extract $\mu_B/T$, $\mu_S/T$ and $\mu_Q/T$. We use experimentally measured particle–antiparticle specific  ratios 
 for proton $p$, Lambda $\Lambda$, and cascade $\Xi$, 
for $\sqrt{s_{NN}}=$ 7.7--39\,GeV from RHIC BES phase-1 to extract the $\mu_{B,S,Q}/T$. These compared well with published STAR freeze-out parameters. These combinations are verified to predict a similar combination of $\Omega$ yields. We also extend this idea to predict (anti-)nuclei yields at energies where they are not measured. We also update parametrizations for the $\sqrt{s_{NN}}$ dependence of freeze-out parameters $T$ and $\mu_B$, and present for the first time a similar parametrization of $\mu_S$. 
}
\maketitle
\baselineskip=\normalbaselineskip

\section{Introduction}

Relativistic heavy-ion collisions create hot and dense QCD matter whose 
properties and phase structure are believed to be encoded in final-state 
hadron abundances. Chemical freeze-out refers to the stage where inelastic 
collisions cease and particle yields are fixed. 
The statistical hadronization model (SHM) provides a successful framework to describe particle production in relativistic heavy-ion collisions by assuming that, at chemical freeze-out, the system attains thermal and chemical equilibrium. In this approach, the yields of identified hadrons are determined by a small set of thermodynamic parameters,
mainly the temperature and chemical potentials ($T$, $\mu_B$,
$\mu_S$, $\mu_Q$). 
The underlying description is often based on the hadron resonance gas (HRG) model, which treats the hadronic phase as a non-interacting gas of all known hadrons and resonances and has been successful in reproducing bulk thermodynamic properties of QCD matter \cite{
Hagedorn:1980kb,Hagedorn:1984uy,Becattini:2016xct,Bazavov:2017dus,
Chatterjee:2013yga}.
By fitting measured particle yields or their ratios, one can extract the chemical freeze-out conditions and gain insight into the QCD phase structure. These fits depend on hadron lists, decay treatment, and ensemble choice. The SHM has been widely applied across different collision energies, demonstrating a remarkable ability to reproduce experimental data over a broad range of hadron species 
~\cite{Becattini:2003wp,Manninen:2008mg,Becattini:2010sk,Andronic_2018,Braun-Munzinger:2003pwq,Andronic:2005yp,Braun-Munzinger:2015hba,STAR:2017sal, Sharma:2018jqf, Sharma:2018owb,Braun-Munzinger:2024ybd}.

An approach we advocate here is to extract parameter ratios $\mu_{B,Q,S}/T$ 
through experimentally measured particle–antiparticle ratios. These ratios of 
observables do not involve an overall volume factor nor any masses. They 
lead to relations that can be analytically inverted, and thereby reduce 
certain systematic uncertainties. In this paper, we implement this technique 
on published hadron yields and compare the extracted chemical potential ratios 
with the results of thermal-model fit obtained by STAR \cite{STAR:2017sal} 
using THERMUS \cite{Wheaton:2004qb}. This approach can also be extended to 
yields of light nuclei and anti-nuclei. These are the main results presented 
in this paper.

The paper is organized as follows. Section \ref{sec:theory} summarizes
the analytic relations used and their inversion. Section \ref{sec:data} 
describes the data and the analysis choices. A simple test of the approach to 
thermalization is presented in Section \ref{sec:doubleRatio}. The extracted 
freeze-out parameters, namely $\mu_{B,Q,S}/T$, their centrality and energy 
dependence, and an independent verification, are given in Section 
\ref{sec:results}. In Section \ref{sec:nuclei} the method is extended to light 
(anti-)nuclei, and implications for fluctuation observables are discussed. 
Conclusions are presented in Section \ref{sec:conclusions}.

\section{\label{sec:theory}Using the statistical thermal model}

The statistical thermal model assumes that all hadrons  follow equilibrium
distributions at freeze-out in a high energy collision. Hadron abundances are fixed at chemical freeze-out, defined to be the thermodynamic conditions at which inelastic collisions cease. while the particle spectral distribution offers insight into the kinetic freeze-out conditions where elastic collisions cease. In the grand-canonical (GC) ensemble, the conserved quantities of the system, namely the baryon number $B$, the strangeness $S$, and the charge $Q$, are described by the respective chemical potentials $\mu_B$, $\mu_S$ and $\mu_Q$. A fugacity $\lambda_{B,Q,S}=\exp(\mu_{B,Q,S}/T)$ can be defined for each chemical potential, where $T$ is the temperature of the system. The ensemble-averaged density of hadron species $i$ with mass $m_i$ and quantum numbers $B_i$, $S_i$ and $Q_i$ is
\begin{equation}
n_i(T,\mu_B,\mu_S,\mu_Q) =
\frac{g_i}{2\pi^2}m_i^2T\lambda_B^{B_i}\lambda_S^{S_i}\lambda_Q^{Q_i}
K_2\left(\frac{m_i}T\right)
\label{eq:dens},
\end{equation}
where, $g_i$ denotes the spin-isospin degeneracy factor and $K_{2}(x)$ is the
second order modified Bessel function. The factor $K_2$ is obtained for the
particles $i$ for which $m_i/T \gg 1$, so the Bose and Fermi distributions are well approximated by the Boltzmann distribution. Practically, at freeze-out this is true for all particles except pions. In using these to fit experimentally measured yields, the condition of strangeness neutrality and fixed baryon-to-charge ratio in the initial state are typically used to constrain the chemical potentials $\mu_S$ and $\mu_Q$.

For a given hadron species $h$, the antiparticle $\bar h$, has quantum 
numbers $B_{\bar h}=-B_h$, $Q_{\bar h}=-Q_h$ and $S_{\bar h}=-S_h$. Eq.\ 
(\ref{eq:dens}) then gives the ratio of particle and anti-particle
densities as a simple function
\begin{eqnarray}
\nonumber
\frac{n_{h}}{n_{\rm\overline{h}}} 
&=&  \lambda_B^{2B_h}\lambda_S^{2S_h}\lambda_Q^{2Q_h}\\
&=& \exp\left[\frac{2 (B_h \mu_B 
+ S_h \mu_S + Q_h \mu_Q)}T\right].
\label{eq:pbarp}\end{eqnarray}
This is computationally simpler than Eq.\ (\ref{eq:dens}), since the complicated dependence on the mass cancels out.

Some double ratios are even simpler. For example, we find that
\begin{equation}
\frac{\left(n_\Lambda / n_{\overline{\Lambda}} \right)}
{\left(n_p / n_{\overline{p}} \right)}
\approx \frac{\left( n_\Xi / n_{\overline{\Xi}} \right)}
{\left( n_\Lambda / n_{\overline{\Lambda}} \right)} 
\approx \frac{n_{K^-}}{n_{K^+}} = \exp\left[\frac{-2\mu_S}T\right],
\label{eq:lpxk}
\end{equation}
since $\mu_B/T$ cancels between the numerator and denominator in the first two, and the approximation involved is that $\mu_Q/T$ is small. Such (approximate) equalities of double ratios are checks of local thermal equilibrium that are easily performed before fitting.
Throughout this work, we assume simultaneous freeze-out of all particles, implying a common freeze-out temperature.

Higher cumulants of the distribution probe thermalization even more 
intensively \cite{Mark:2024hrv}. These higher order susceptibilities, $\chi^n$ 
(for $n\ge2$), are constructed experimentally by examining the fluctuations of 
individual baryons, which we can denote as $\chi^n_h$. In the Boltzmann limit, 
relations similar to Eq.\ (\ref{eq:lpxk}) hold for all these $\chi^n_h$. One 
finds
\begin{equation}
\frac{\left(\chi^n_\Lambda / \chi^n_{\overline{\Lambda}} \right)}
{\left(\chi^n_p / \chi^n_{\overline{p}} \right)}
\approx \frac{\left(\chi^n_\Xi / \chi^n_{\overline{\Xi}} \right)}
{\left(\chi^n_\Lambda / \chi^n_{\overline{\Lambda}} \right)} 
\approx \frac{n_{K^-}}{n_{K^+}}.
\label{eq:cumulants}\end{equation}
 Higher cumulants of the baryon distribution were found to raise questions about complete thermalization at lower energies in BES-I \cite{Gupta:2022phu}. In view of this, we anticipate that tests such as those in Eq.\ (\ref{eq:cumulants}) will become very important for forthcoming experiments with high intensity and very low beam energies, where $\chi^n_\Lambda$ and $\chi^n_\Xi$ may be measured. We emphasize that the double ratios of cumulants of all orders are equal to the ratio $n_{K^-}/n_{K^+}$ in the Boltzmann limit.

Using Eq.\ (\ref{eq:dens}) one finds that the ratio ($R_h$) of the 
arithmetic and geometric means of $n_h$ and $n_{\bar h}$
\begin{eqnarray}
\nonumber
R_h &\equiv& \frac{n_h + n_{\bar{h}}}{2\sqrt{n_h n_{\bar{h}}}} = \frac12\left(\sqrt{\frac{n_h}{n_{\bar h}}}+\sqrt{\frac{n_{\bar h}}{n_h}}\right)\\
&=& \cosh\left(\frac{B_h\mu_B + S_h\mu_S + Q_h\mu_Q}{T}\right).
\label{eq:single_ratio}
\end{eqnarray}
The quantity $R_h$ depends only on the quantum numbers $B_h$, $S_h$, $Q_h$ and the chemical potentials, independent of the mass and degeneracy factors. 
For protons $p$, Lambda $\Lambda$, cascade $\Xi$, and Omega $\Omega$ baryons one can write
\begin{eqnarray}
R_p &=& \cosh\left(\frac{\mu_B + \mu_Q}{T}\right), \label{eq:P} \\
R_{\Lambda} &=& \cosh\left(\frac{\mu_B - \mu_S}{T}\right), \\
R_{\Xi} &=& \cosh\left(\frac{\mu_B - 2\mu_S + \mu_Q}{T}\right), \label{eq:X}\\
R_{\Omega} &=& \cosh\left(\frac{\mu_B - 3\mu_S + \mu_Q}{T}\right). 
\label{eq:observables}
\end{eqnarray}
Any three of these can be inverted to find the three $\mu/T$ ratios. We
choose to use Eqs.\ (\ref{eq:P}--\ref{eq:X}) along with experimentally
measured yields of $p$, $\Lambda$ and $\Xi$, and leave $R_\Omega$ as a
check. This choice is prompted by the fact that the yields of $\Omega$ and $\overline\Omega$ are very small because of their larger masses, and hence the measurements come with larger relative uncertainties.

Inverting the above equations gives the expressions
\begin{align}
\frac{\mu_B}{T} &= \frac{1}{2}\left[\cosh^{-1}(R_p)  - \cosh^{-1}(R_{\Xi})\right] + \cosh^{-1}(R_{\Lambda}), \label{eq:x}\\
\frac{\mu_Q}{T} &= \frac{1}{2}\left[\cosh^{-1}(R_p) + \cosh^{-1}(R_{\Xi})\right] - \cosh^{-1}(R_{\Lambda}), \label{eq:y} \\
\frac{\mu_S}{T} &= \frac{1}{2}\left[\cosh^{-1}(R_p) - \cosh^{-1}(R_{\Xi})\right]. \label{eq:z}
\end{align}
The uncertainties in $R_P, R_{\Lambda}, R_{\Xi}$ propagate to $\frac{\mu_B}{T}$, $\frac{\mu_Q}{T}$, and $\frac{\mu_S}{T}$ using standard linear error propagation.

At LHC energies, these fugacities are close to unity, so particles 
and anti-particles are produced in equal abundance and particle to
anti-particle ratios are unity. On the other hand, data from
the STAR collaboration probe a range of values of the fugacities,
and provide us with a non-trivial  opportunity to check the validity
of Eq.\ \ref{eq:lpxk}. In this context, note that in the published
results \cite{STAR:2017sal}, $\mu_Q$ is small and is often taken to be zero.
However, in our method, we retain $\mu_Q$ as an extracted parameter.

\section{Data and analysis details}\label{sec:data}

This work employs identified hadron yields from the STAR for various 
collision energies and centralities~\cite{STAR:2017sal}. For each beam 
energy and centrality bin, we computed $R_p, R_\Lambda, R_\Xi$ using Eqs.\ 
(\ref{eq:P}--\ref{eq:X}) using the published mid-rapidity yields.
For energy dependence, the published particle yields from 
AGS~\cite{E-802:1998xum, Ahmad:1991nv, Albergo:2002tn, Ahmad:1998sg, E917:2001eko} and SPS~\cite{NA49:2006gaj, NA49:2008ysv} experiments are also
used.

Key systematic considerations include feed-down corrections (particularly for
protons), acceptance and efficiency differences between particles and 
antiparticles. Double ratios mitigate some, but not all, of these 
uncertainties. At low energy, baryon stopping changes the composition at mid-rapidity. This must be treated carefully when comparing to global fits. Most 
of these considerations have been taken into account in the results published 
by the experimental collaborations. However, corrections due to the weak-decay 
feed-down for protons still need to be performed. 
For STAR's $p$ and $\bar p$ these corrections were obtained using the THERMUS code~\cite{Wheaton:2004qb}.  The ratios are
computed using central values of measured yields, while the uncertainties on
the ratios are evaluated by standard error propagation assuming uncorrelated
errors.

\section{Test of thermalization}\label{sec:doubleRatio}

\begin{figure}[!ht]
\centering
\includegraphics[width=0.45\textwidth]{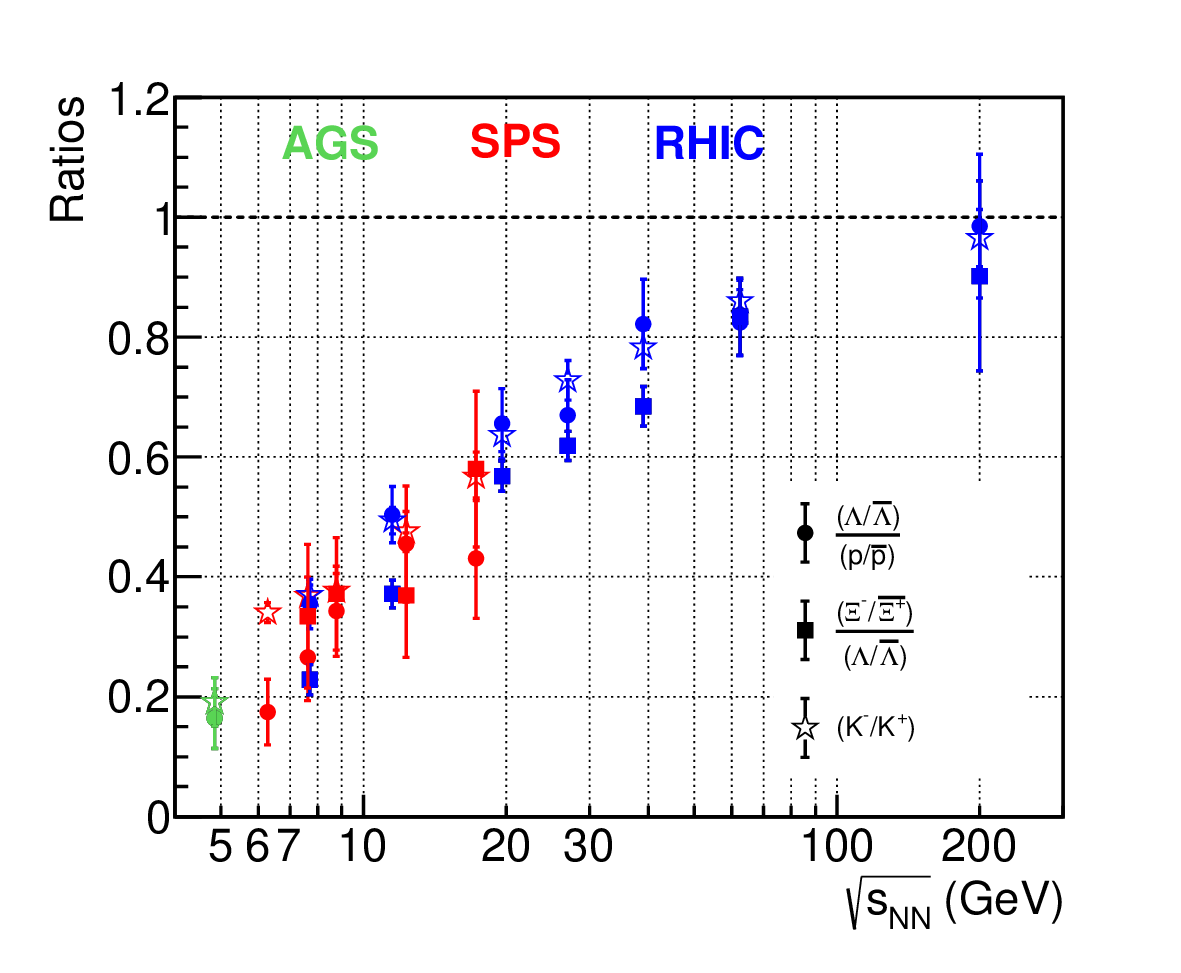}
  \caption{Double ratios of various particles as given in Eq.\ 
  (\ref{eq:lpxk}) displayed as functions of the energy $\sqrt{s_{NN}}$.}
\label{fig:RatioVsEnergy}
\end{figure}

We start by testing the validity of a statistical hadronization model using 
Eq.\ (\ref{eq:lpxk}) in the various collision energies studied. As we discussed
in Section \ref{sec:theory}, this gives a test of the applicability of the 
model which does not depend on the specific values of its parameters, except that $\mu_Q/T$ is small. The ratios are constructed using experimentally measured particle yields in AGS \cite{E-802:1998xum, Ahmad:1991nv, Albergo:2002tn, Ahmad:1998sg, E917:2001eko}, SPS \cite{NA49:2006gaj, NA49:2008ysv} and RHIC \cite{STAR:2017sal, Abelev:2008ab}. With these input data, we constructed each of the double ratios given in Eq.\ (\ref{eq:lpxk}).

Figure \ref{fig:RatioVsEnergy} shows the double ratios constructed using $p$, $\Lambda$, $\Xi$, and $K$ as functions of energy. All of these ratios are consistent with each other within uncertainties at higher energy. However, there could be some deviations from equality at lower $\sqrt{s_{NN}}$. With this caveat, the data suggest that, to a good approximation, these particles are thermally produced in high energy heavy-ion collisions with negligible $\mu_Q/T$. The shape of the curve is a reflection of the change of $\mu_S/T$ with $\sqrt{s_{NN}}$. At very high energies ($\sqrt{s_{NN}} \geq 200$ GeV), all the ratios approach unity, reflecting the fact that $\mu_S/T$ becomes small.

The ratios presented here include uncertainties obtained by standard error propagation of the published experimental yields. This could lead to overestimates of the uncertainties shown in Figure \ref{fig:RatioVsEnergy}. In future experiments or data analyses, individual sources of systematic uncertainty could be minimized by directly estimating the experimental uncertainties on the particle-antiparticle ratios. This would lead to the cancellation of correlated systematic effects to a large extent. A more precise estimation of these uncertainties would provide stringent tests of thermalization through the double ratios, especially in future high statistics experiments at low energy, starting from BES-II.

\section{Results}\label{sec:results}

\subsection{Centrality dependence of chemical potential ratios}\label{sec:newfits}
\begin{figure}[!ht]
\centering
\includegraphics[width=0.40\textwidth]{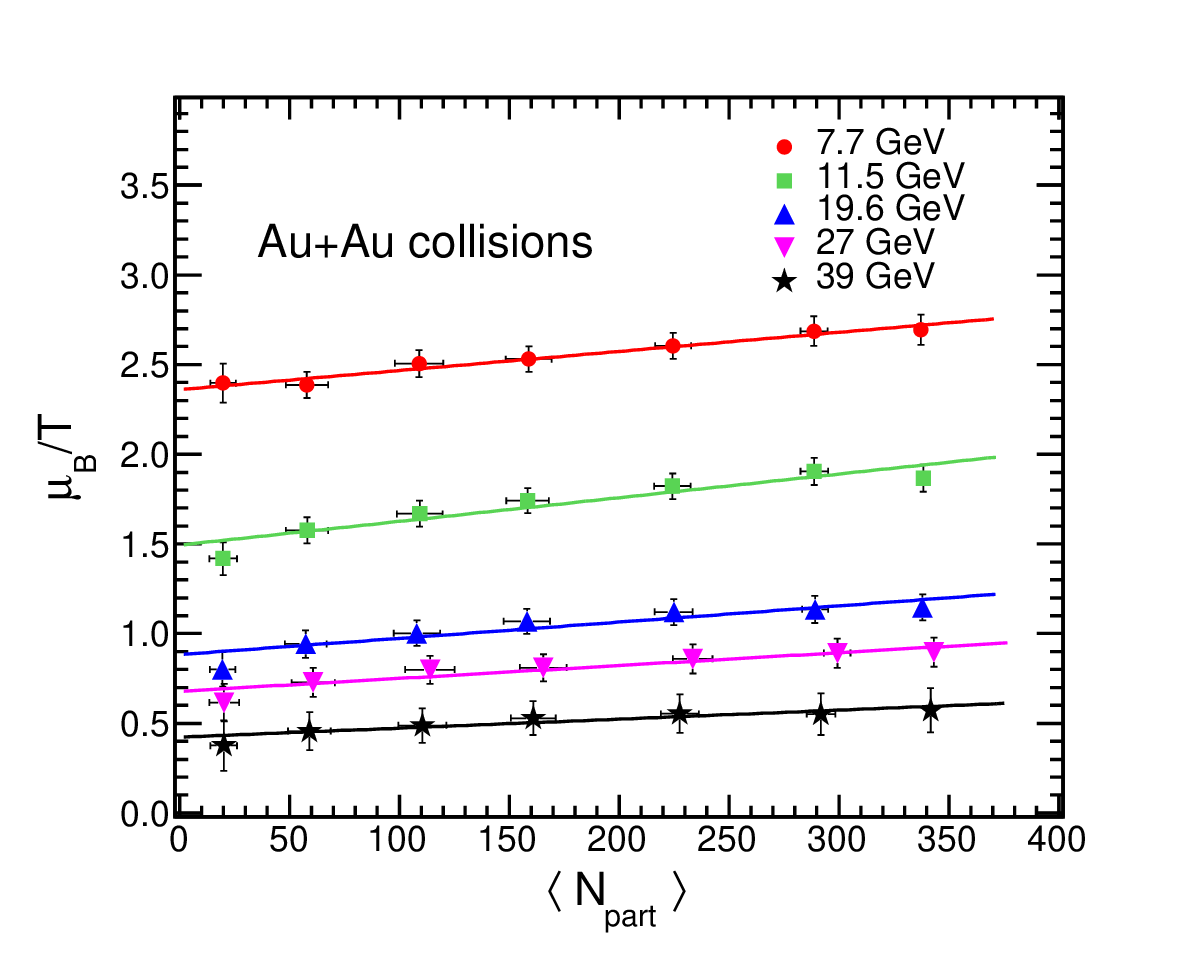}
\includegraphics[width=0.40\textwidth]{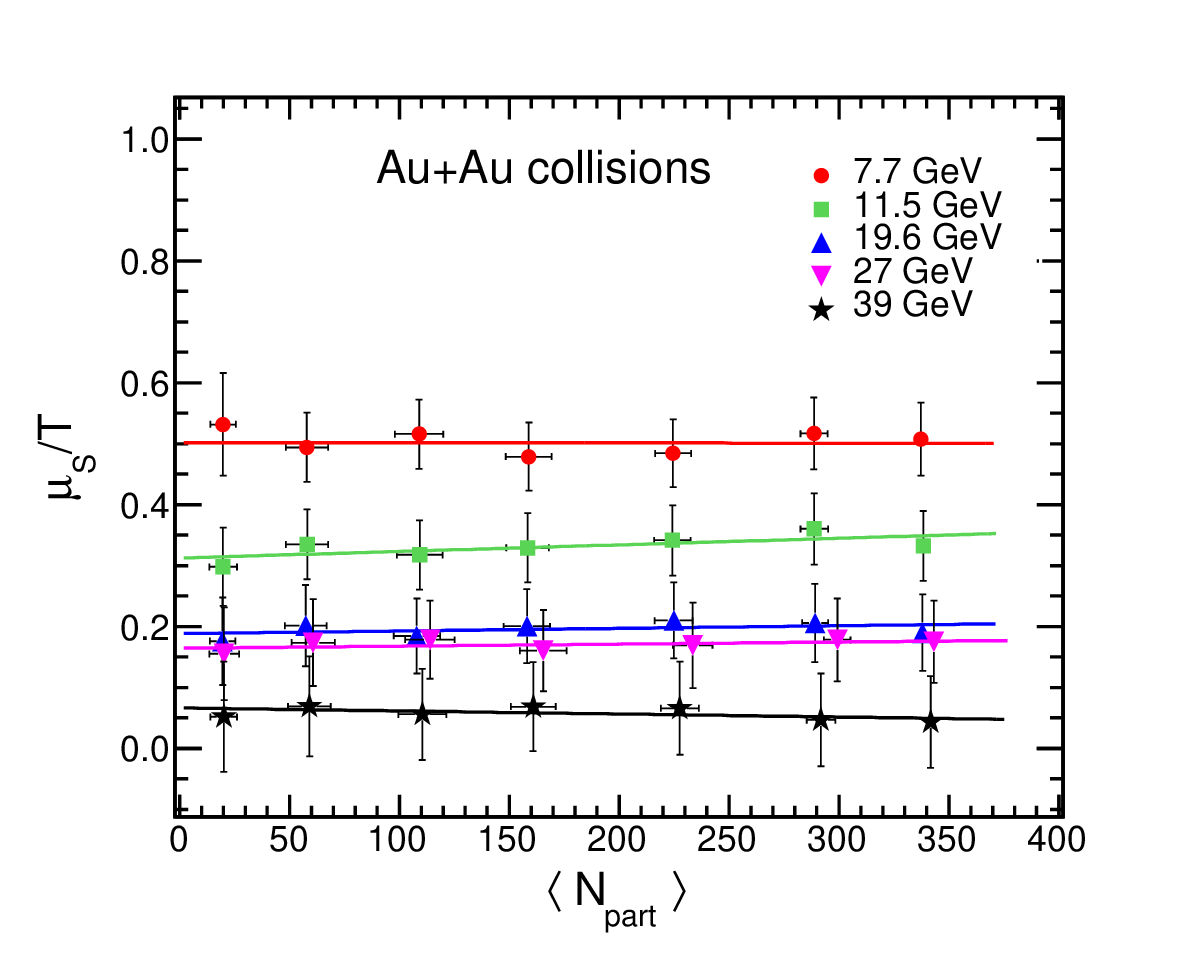}
\includegraphics[width=0.40\textwidth]{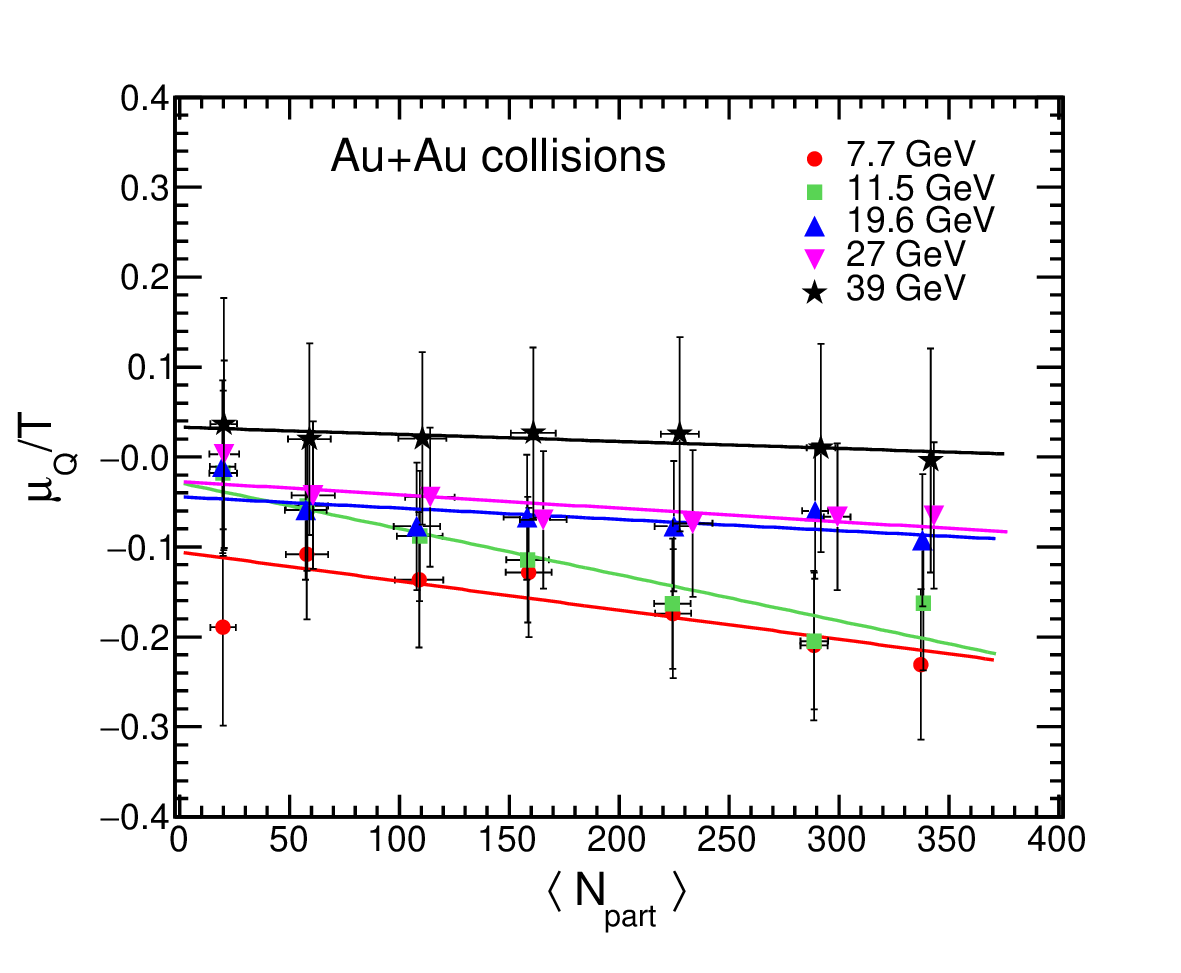}
t\caption{Extracted $\mu_B/T$, $\mu_S/T$, and $\mu_Q/T$  ratios as a
  function of number of participating nucleons $\langle N_{\rm{part}}
  \rangle$ in various energies. 
  Lines represent the linear fits to the data points.}
	\label{fig:param}
\end{figure}
\begin{figure}[!ht]
\centering
\includegraphics[width=0.35\textwidth]{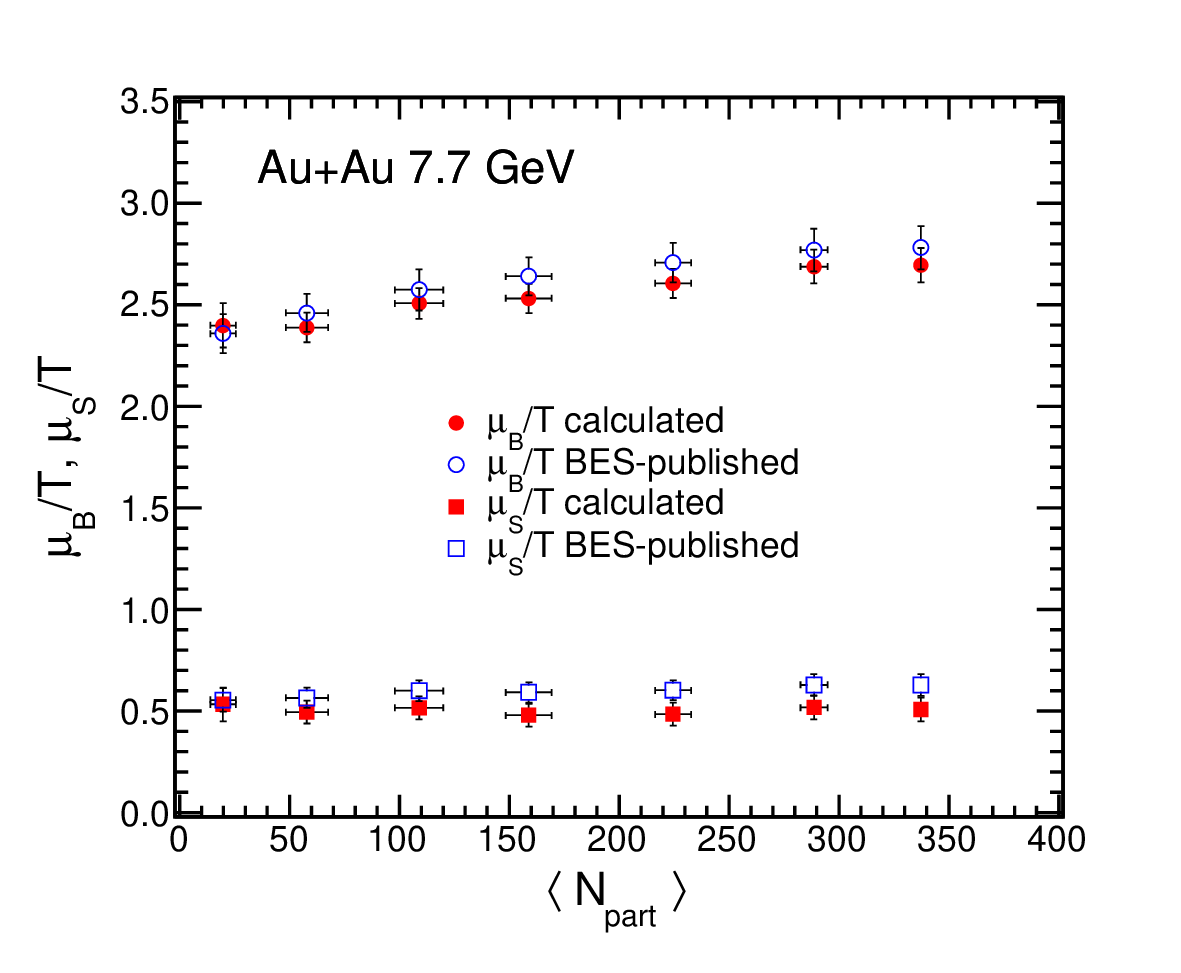}
\includegraphics[width=0.35\textwidth]{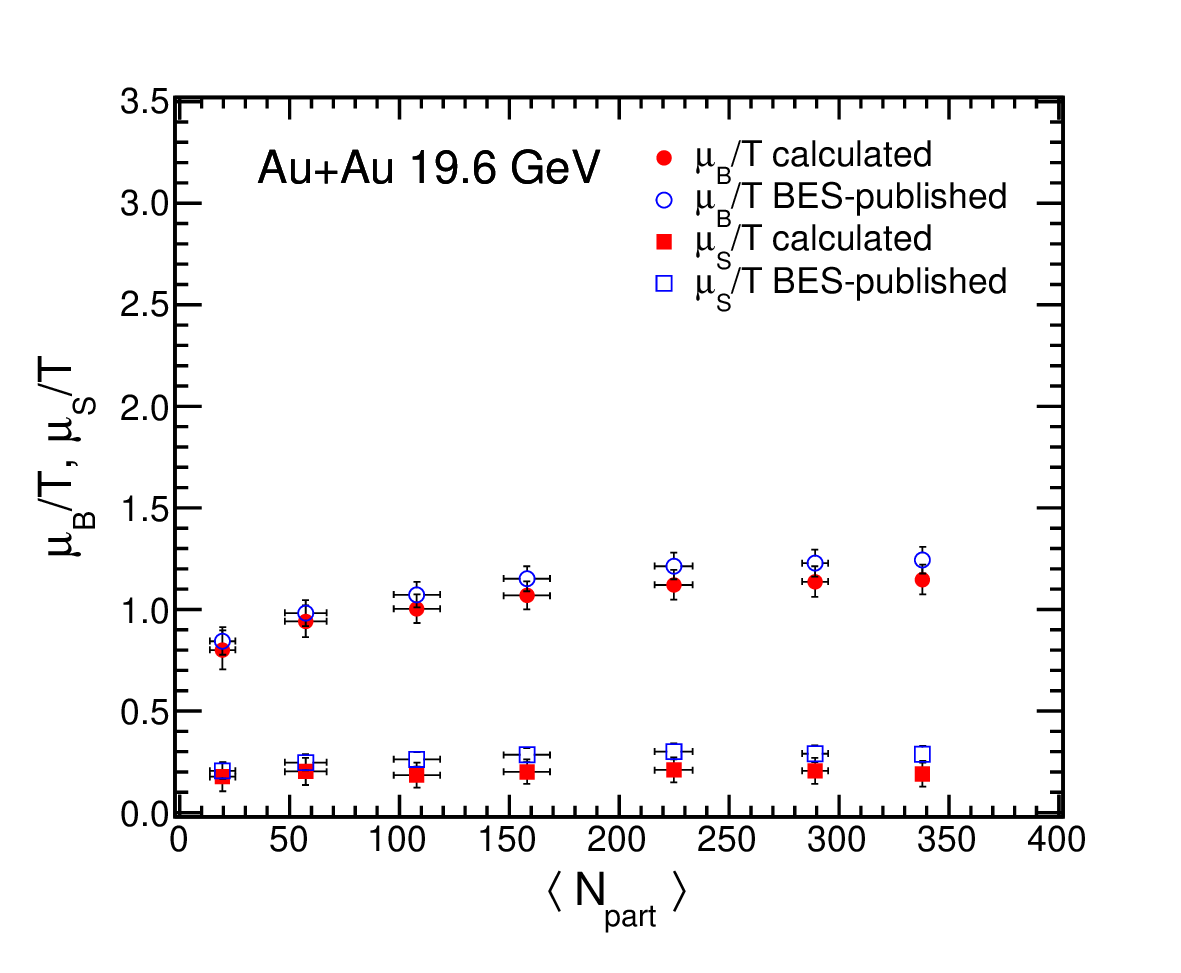}
\includegraphics[width=0.35\textwidth]{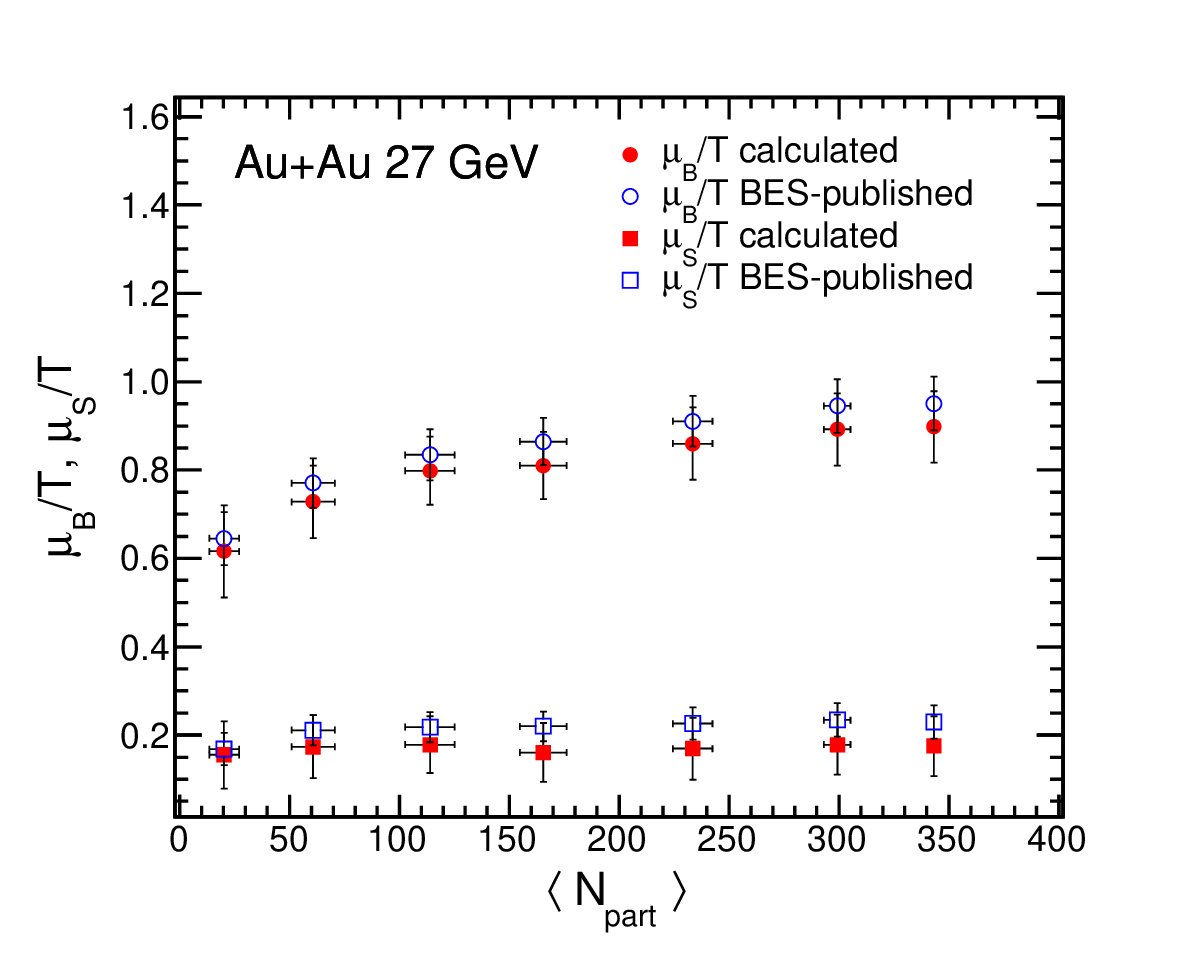}
\includegraphics[width=0.35\textwidth]{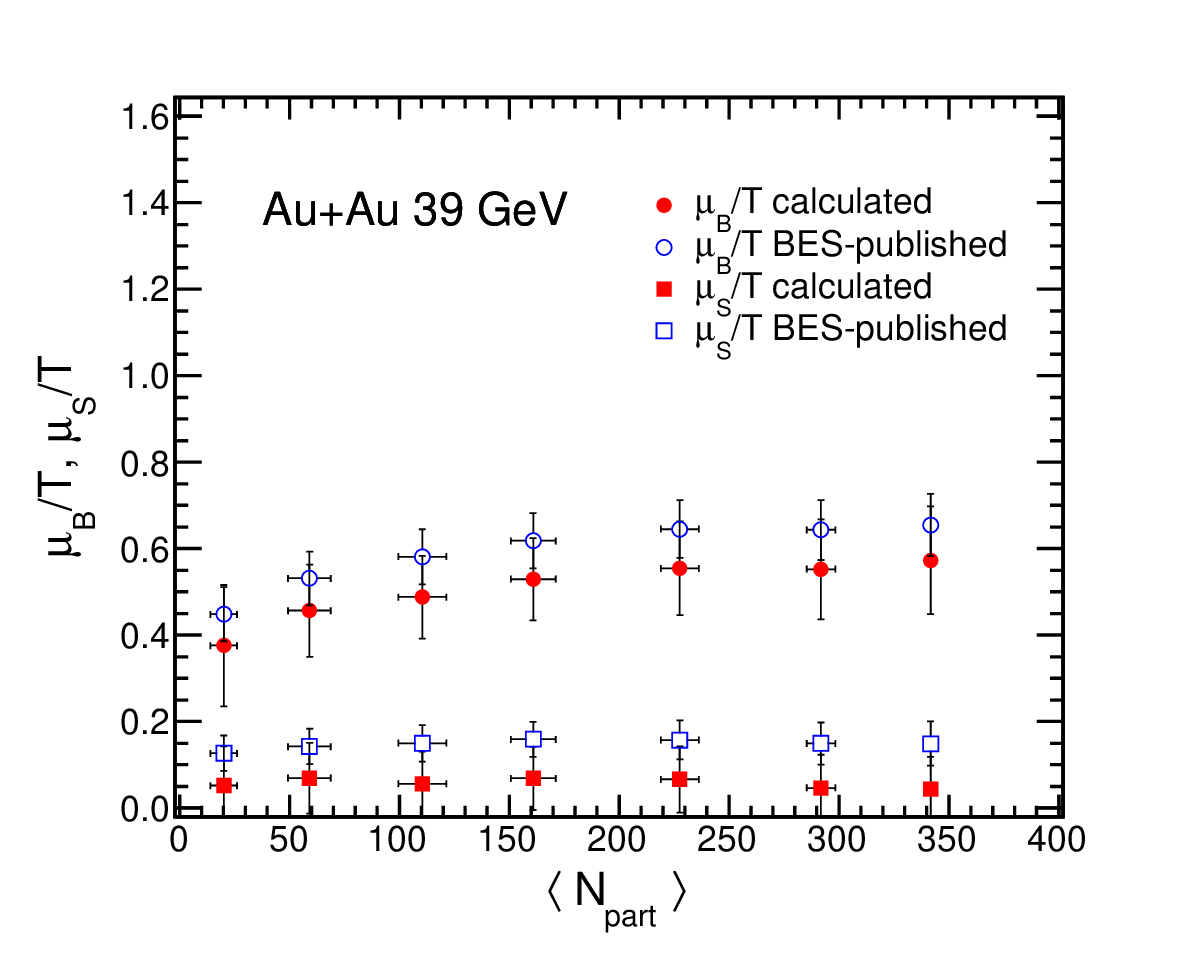}
  \caption{Extracted $\mu_B/T$ and $\mu_S/T$ ratios at different
    energies and their comparison with the published freeze-out
    parameters~\cite{STAR:2017sal}.}
	\label{fig:param_comp}
      \end{figure}

We have extracted $\mu_{B,S,Q}/T$ ratios at chemical freeze-out as functions
of number of participant nucleons ($\langle N_{\rm{part}}  \rangle$) using
Eqs.\ (\ref{eq:x})--(\ref{eq:z}) for gold-on-gold (Au+Au) collisions at
$\sqrt{s_{NN}}$ of 7.7, 11.5, 19.6, 27 and 39 GeV. The results obtained are
shown in Figure \ref{fig:param}.

The fits show that the $\mu_B/T$ ratio increases with $\langle N_{\rm{part}}
\rangle$. This correctly reflects increasing baryon density in more central 
collisions. The consistent decrease in this ratio with increasing beam
energy is consistent with the idea that baryon stopping decreases with
increasing collision energy. The $\mu_S/T$ ratio is independent of $\langle 
N_{\rm{part}} \rangle$, as expected, since the participants have a fixed 
strangeness. Observe the significant decrease of $\mu_s/T$ with beam energy, 
where more strange particles are expected to be created at higher energy. This 
is consistent with the OZI rule, which implies that strange and anti-strange
particles are created together. The quantity $\mu_Q/T$ is small and therefore
has large relative errors, so any statement about trends is necessarily weak.

The smallness of $\mu_Q/T$ is related to the fact that the systems
are nearly isoscalar. It has a tendency to become more negative with 
increasing $\langle  N_{\rm{part}} \rangle$. The Gell-Mann-Nishijima
relation, $Q=I_3+(B+S)/2$ allows us to understand this trend, since
increasing the number of participants involves increasing the number of 
baryons in the system. Similarly, the observation that $\mu_Q/T$ moves 
closer to zero with increasing beam energy has to do with the fact that 
baryon stopping decreases with energy. In the future, it may be interesting
to extract the chemical potential associated with the isospin component
$I_3$. This is also the parameter used in lattice computations, so this
change may be useful direct in comparisons with QCD.

In Figure \ref{fig:param_comp} we show the $\mu_B/T$  and $\mu_S/T$ ratios
as a function of $\langle N_{\rm{part}} \rangle$ at the four energies 7.7,
19.6, 27 and 39 GeV. The open symbols represent the published freeze-out parameter ratios~\cite{STAR:2017sal}, while the solid symbols show the
results of this fit. The good agreement between our results and the
published fits from the STAR experiment is a validation of the
method used here. Due to the larger relative errors on $\mu_Q/T$, its
comparison does not add anything to this conclusion.

\subsection{Validation using $R_\Omega$}

\begin{figure}[!htb]
\centering
\includegraphics[width=0.45\textwidth]
{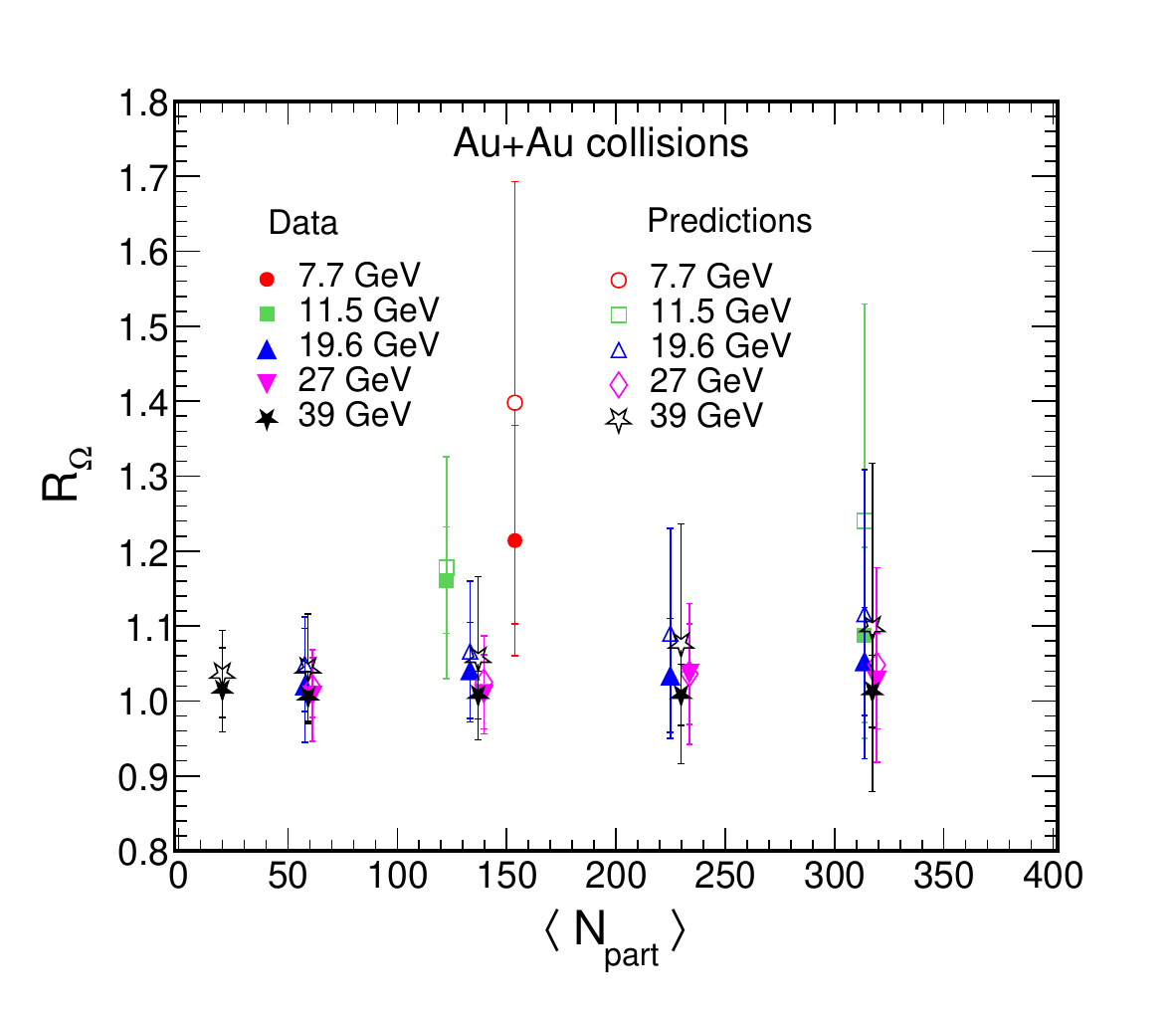}
  \caption{Verification of $R_\Omega$ predicted by Eq.\ (\ref{eq:observables}).
  The quantity is computed from experimentally measured $\Omega$ and
 $\overline{\Omega}$ yields, while the $\cosh$ is obtained using the  chemical  potential ratios obtained using Eqs.\ (\ref{eq:P}--\ref{eq:X}).}
  \label{fig:omega}
\end{figure}

As mentioned above, $R_{\Omega}$ provides a validation of the method.
The experimentally 
measured $\Omega$ and $\overline{\Omega}$  yields
 are used
to estimate $R_{\Omega}$. Using the extracted values of $\mu_B/T, \mu_S$/T and
$\mu_Q/T$ one obtains model predictions for this quantity using Eq.\ 
(\ref{eq:observables}).  The uncertainties in $R_{\Omega}$ (data) are evaluated using the standard error propagation from the uncertainties in the measured hadron yields, as described in Eq.\ \ref{eq:single_ratio}. For the predicted $R_{\Omega}$, the uncertainties are obtained by propagating the uncertainties in the estimated $\mu_B/T, \mu_S$/T and $\mu_Q/T$.

Figure \ref{fig:omega}
shows this comparison by plotting $R_\Omega$ obtained from experiments along
with the value predicted by Eq.\ (\ref{eq:observables}). The two are shown
as functions of $\langle N_{\rm{part}} \rangle$ in Au+Au
collisions at $\sqrt{s_{NN}} = $ 7.7, 11.5, 19.6, 27 and 39 GeV. The
complete consistency of the measurements and predictions is a nontrivial
validation of Eq.\ (\ref{eq:observables}). This test is even more
interesting since $\Omega$ is a multi-strange baryon and so is more
sensitive to $\mu_S/T$ than any of the inputs.

\subsection{Energy dependence of $R_p$, $R_\Lambda$ and $R_\Xi$ \label{edep_dr}}

\begin{figure}[!hbt]
\centering
\includegraphics[width=0.45\textwidth]
{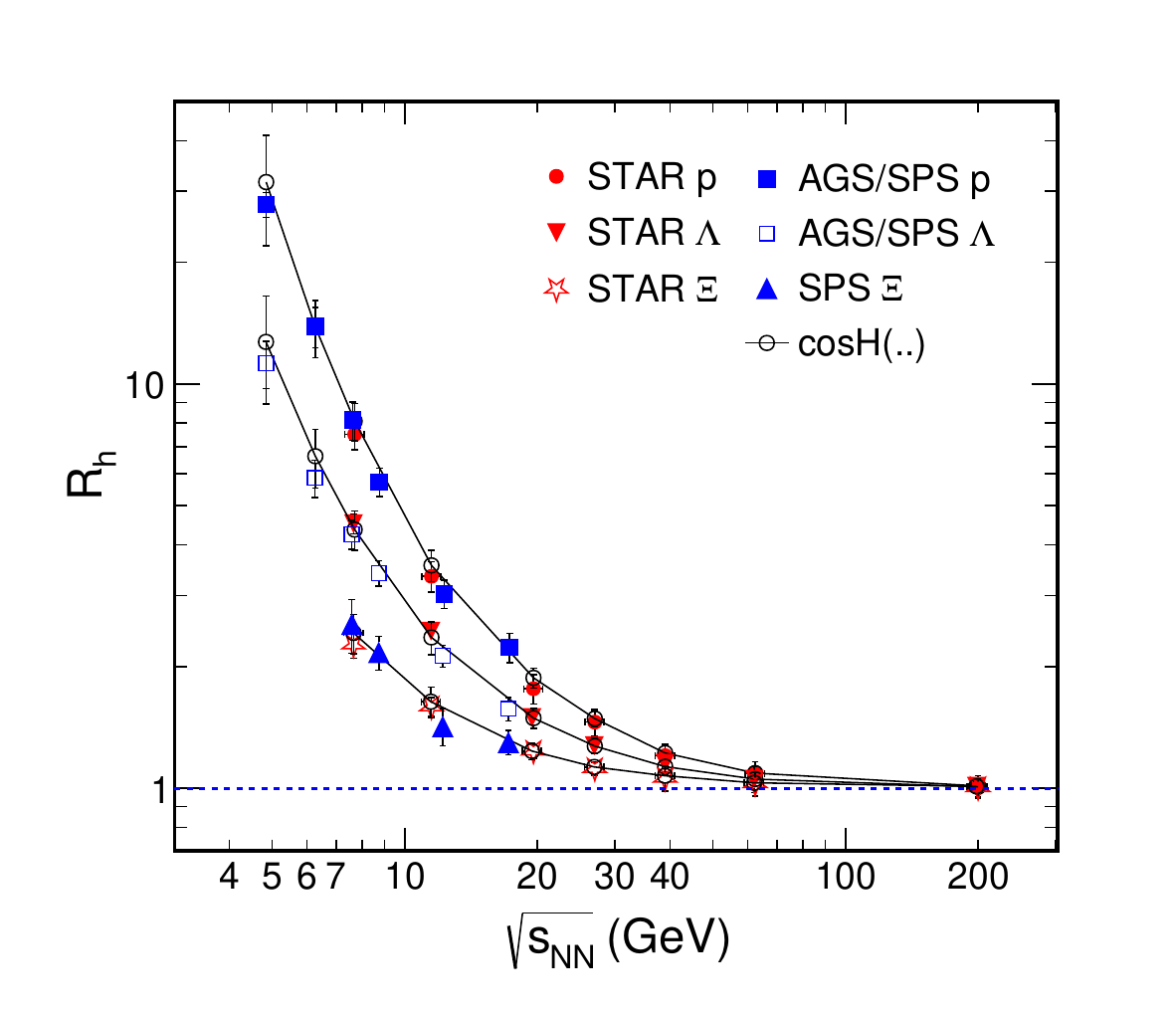}
  \caption{Energy dependence of ratios $R_p$, $R_\Lambda$, and $R_\Xi$ 
  obtained from published hadron yields \cite{E-802:1998xum, Ahmad:1991nv,
  Albergo:2002tn, Ahmad:1998sg, E917:2001eko, NA49:2006gaj, NA49:2008ysv,
  STAR:2017sal, Abelev:2008ab}. Measurements (plotted as points) are
  compared with model predictions (the smooth curves) obtained using
  Eqs.\ (\ref{eq:P}--\ref{eq:X}).}
\label{fig:edep}
\end{figure}

Next, we present the energy dependence of $R_p$, $R_\Lambda$, and $R_\Xi$.
The measured values are taken from AGS \cite{E-802:1998xum, Ahmad:1991nv, 
Albergo:2002tn, Ahmad:1998sg, E917:2001eko}, SPS \cite{NA49:2006gaj, 
NA49:2008ysv} and RHIC~\cite{STAR:2017sal, Abelev:2008ab}. As mentioned 
above, the reported yields of $p$ and $\bar p$ from STAR were corrected 
for feed down using THERMUS. The good continuity with the AGS and SPS data 
shown in Figure \ref{fig:edep} suggests that this correction is reliable.

The figure also shows a comparison of the data with the model predictions
from Eqs.\ (\ref{eq:P}--\ref{eq:X}), as functions of the collision energy. 
The values of $\mu/T$ are extracted from STAR data. So are only available at
the energies where the data was taken. The continuous curves are obtained
using an interpolating formula. We discuss this in the next subsection.

Observe that $p$, $\Lambda$, and $\Xi$ are different at lower 
energies, but move towards the universal value of unity with increasing
energy. This indicates that the ratios $\mu/T$ decrease rapidly with 
increasing energy. The differences between $p$, $\Lambda$, and $\Xi$ can be 
attributed to $\mu_S/T$. This does not enter $R_p$, but for $\Xi$ its effect
is the square of what it is for $\Lambda$. This is visible on the log-log
plot at every collision energy, as the vertical displacement between $R_p$ and 
$R_\Lambda$ is the same as between $R_\Lambda$ and $R_\Xi$.

\subsection{Energy extrapolations of freeze-out parameters}

The beam energy dependence of the freeze-out parameters has been parametrized
before by many authors \cite{cleymans:2005xv, Manninen:2008mg,
cleymans:2011pe, Andronic:2009jd}. We present an update using thermal model
fits from STAR BES-I data at $\sqrt{s_{NN}}=7.7-200$ GeV.  The availability
of inputs over a larger range of $\sqrt{s_{NN}}$ than was available earlier 
helps to better constrain extrapolations. The input freeze-out parameters
were obtained using the yields of $\pi$, $K$, $p$, $\Lambda$, $\Xi$, and
their antiparticles \cite{STAR:2017sal}. The STAR experiment provides good
estimates of the particle and antiparticle yields even at lower energies,
where the yields are low. In contrast, at AGS and SPS energies, yields of
multi-strange hadrons and their antiparticles are seldom reported. As a
result, the energy-dependent fits provide good estimates of the chemical
potentials at lower AGS and SPS energies.
    
\begin{figure}[!ht]
\centering
\includegraphics[width=0.35\textwidth]{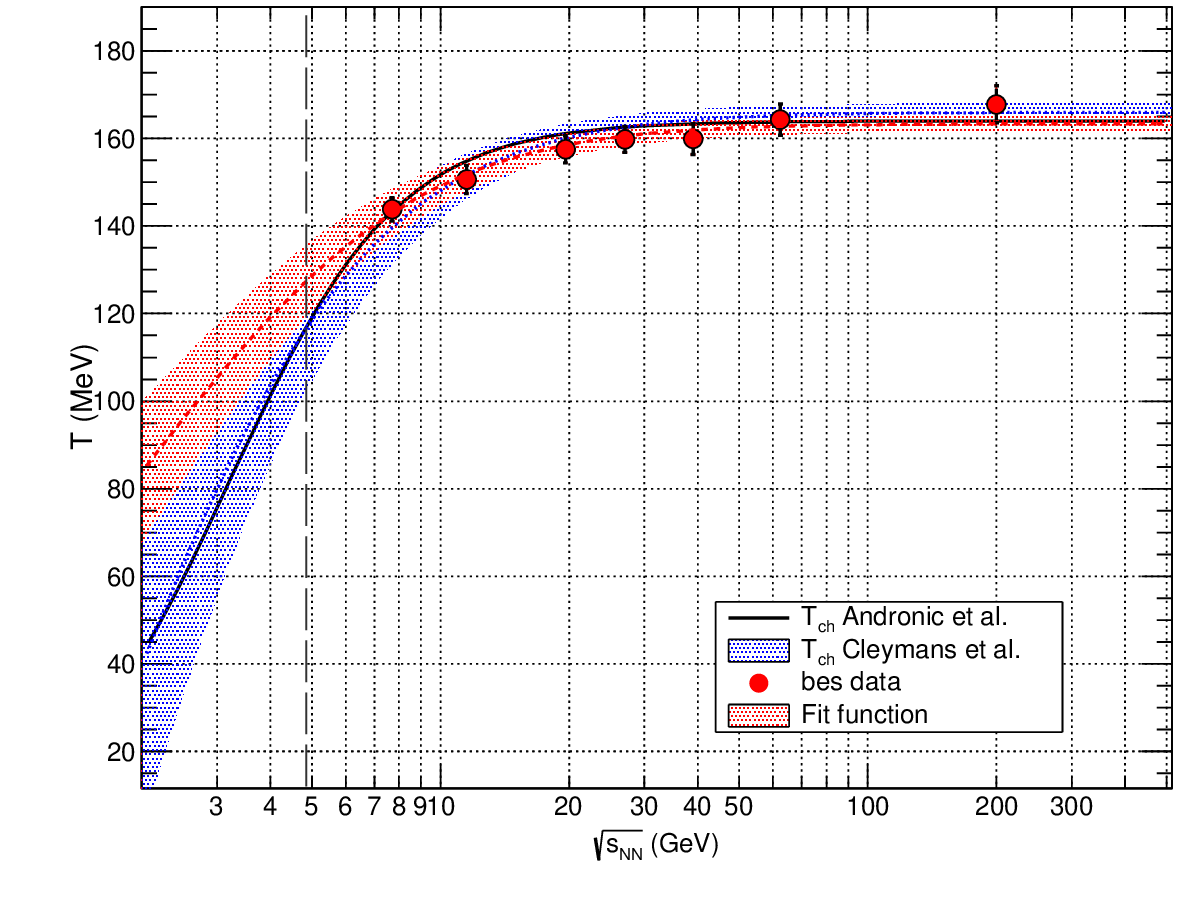}
\includegraphics[width=0.35\textwidth]{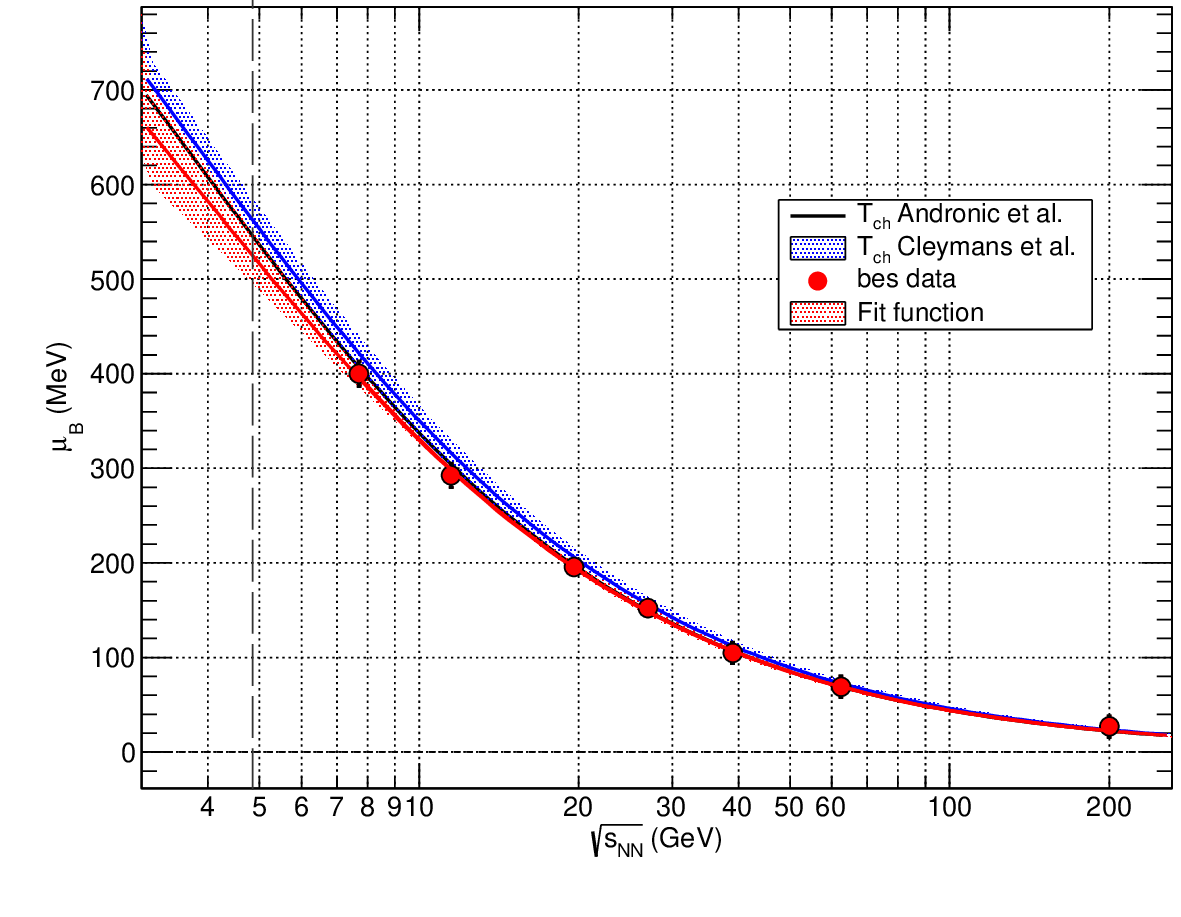}
\includegraphics[width=0.35\textwidth]{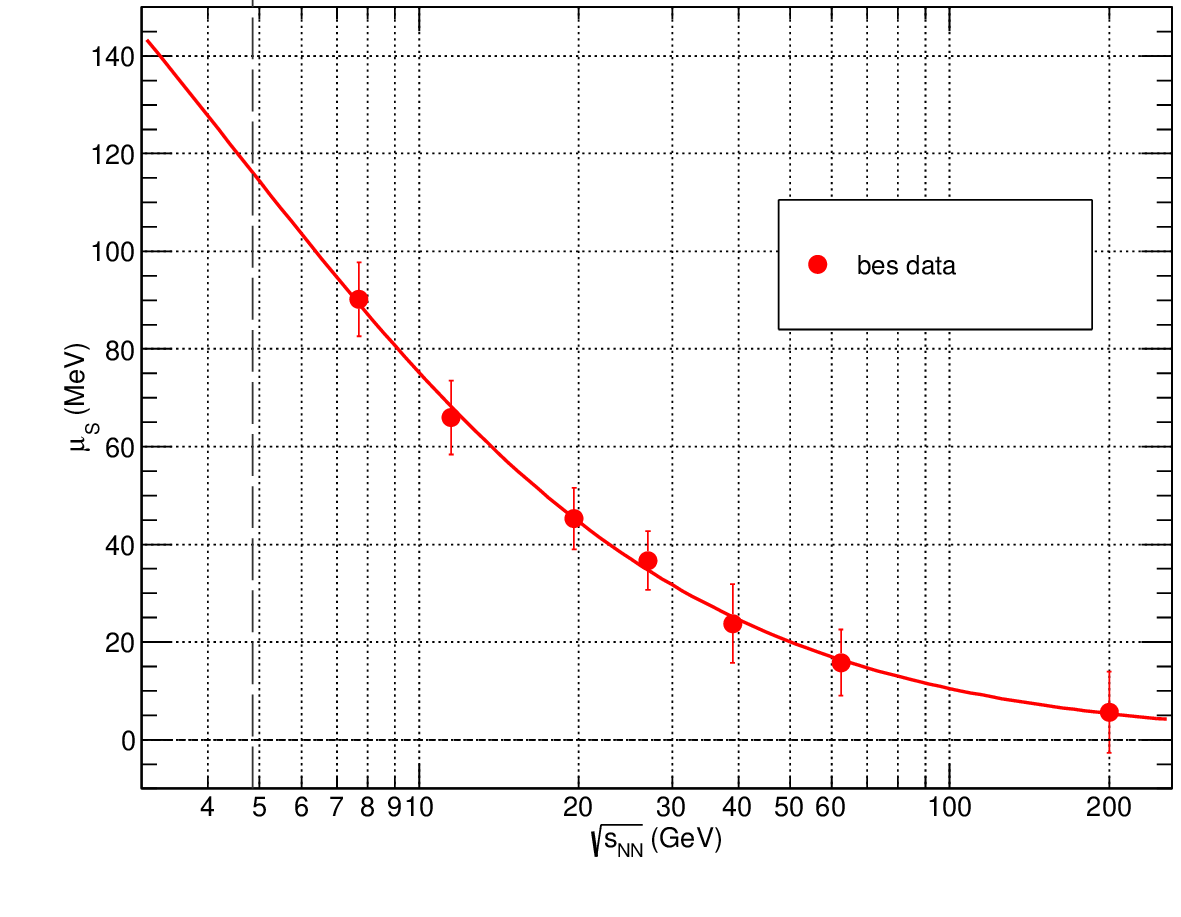}
\includegraphics[width=0.35\textwidth]{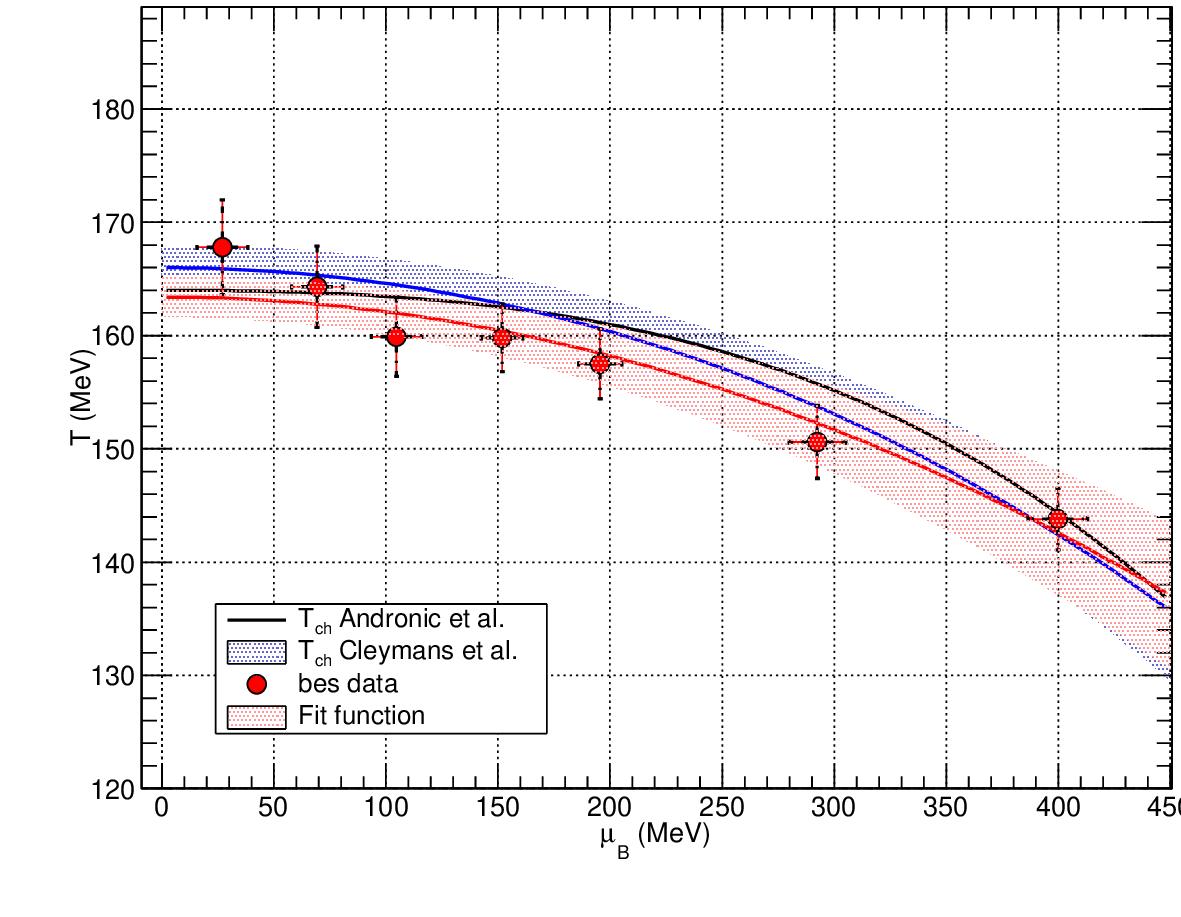}
  \caption{Energy dependence of freeze-out parameters $T$ (top panel),
  $\mu_B$ (second panel), and $\mu_S$ (third panel from top). The panel at
  the bottom shows the freeze-out curve, {\sl i.e.\/}, $T$ plotted against
  $\mu_B$. The gray dashed vertical line at  $\sqrt{s_{NN}} = 4.85$ GeV
  marks the lowest energy for which $T$, $\mu_B$, and $\mu_S$ are extracted
  from data.}
\label{fig:par_edep}
\end{figure}

Figure \ref{fig:par_edep} shows the energy dependence of freeze-out parameters $T$, $\mu_B$, and $\mu_S$, as well as $T_{\rm{ch}}$ versus  $\mu_B$ and fits to them using the following functional forms
\begin{eqnarray}
T &=& A - B\mu_B^2, \label{eq:fit_es_1} \\
\mu_B &=& \frac{C}{1+D\sqrt{s_{NN}}}, \label{eq:fit_ea_2} \\
\mu_S &=& \frac{E}{1+F\sqrt{s_{NN}}}. \label{eq:fit_es_3}
\end{eqnarray}
Using as input the freeze-out parameters from \cite{STAR:2017sal}, we find that 
\begin{eqnarray}
\nonumber
 && A=163.4\pm1.8 {\rm\ MeV\/} \quad B=(1.3\pm0.2)\times10^{-4} \rm{\ MeV\/}^{-1}\\
\nonumber
 && C=1187\pm220 {\rm\ MeV\/} \quad D=0.259\pm0.063 \rm{\ MeV\/}^{-1}\\
 \nonumber
 &&E=239\pm104 {\rm\ MeV\/} \quad F=0.218\pm0.131 \rm{\ MeV\/}^{-1}
\end{eqnarray}
Table \ref{tab:par} shows the extrapolated values at selected energies.

\begin{table}[h!]
  \begin{center}
    \caption{ Extrapolated freeze-out parameters corresponding to collision energies at AGS (4.85 GeV), SPS (6.27 GeV), and RHIC (14.5 GeV).}
    \label{tab:par}
    \begin{tabular}{|c|c|c|c|} 
    \hline 
    \textbf{$\sqrt{s_{NN}}$} (GeV) & \textbf{$T_{\rm{ch}}$} (MeV) & \textbf{$\mu_B$} (MeV) & \textbf{$\mu_S$} (MeV)\\
      \hline
      4.85 & 127.0 $\pm$ 6.5 & 526.4 $\pm$ 27.9 &  116.2 $\pm$ 15.6 \\
      6.27 & 136.4 $\pm$ 3.9 & 452.7 $\pm$ 17.9 & 101.0 $\pm$ 10.2  \\
      14.5 & 154.9 $\pm$ 1.8 & 249.8 $\pm$ 5.7  & 57.5 $\pm$ 3.5\\
      \hline
    \end{tabular}
  \end{center}
\end{table}

For $\sqrt{s_{NN}}=14.5$ GeV STAR has taken data \cite{STAR:2025xxf}, but the freeze-out parameters are not yet published, so the parameters given here are reported as the first estimation of chemical freeze-out parameters at this energy. In addition, the energy dependence of $\mu_S$ is reported first time along with its functional form and fit parameters.

\section{Extension to light nuclei and anti-nuclei prediction}\label{sec:nuclei}

The production mechanism of light nuclei like deuterons ($d$), tritons ($t$),
$^3$He and the corresponding anti-nuclei in high energy collisions is not 
fully understood. Coalescence models suggest that nuclei and anti-nuclei form
when the constituent nucleons lie close in phase space in the final stages of 
evolution of the system. This formation mechanism is also justified by the 
small binding energy ($\sim$ few MeV) of (anti-)nuclei compared to the 
freeze-out temperature ($\sim 150$ MeV) of the system. If the coalescing 
nucleons are thermal and the coalescence process does not change momenta 
appreciably, then their experimentally measured densities ratio  should satisfy
Eq.\ (\ref{eq:single_ratio}) derived from the simple approximation of the 
statistical hadronization model 
\cite{cleymans:2011pe,Sharma:2022poi}.

\begin{figure}[!ht]
\centering
\includegraphics[width=0.45\textwidth]
{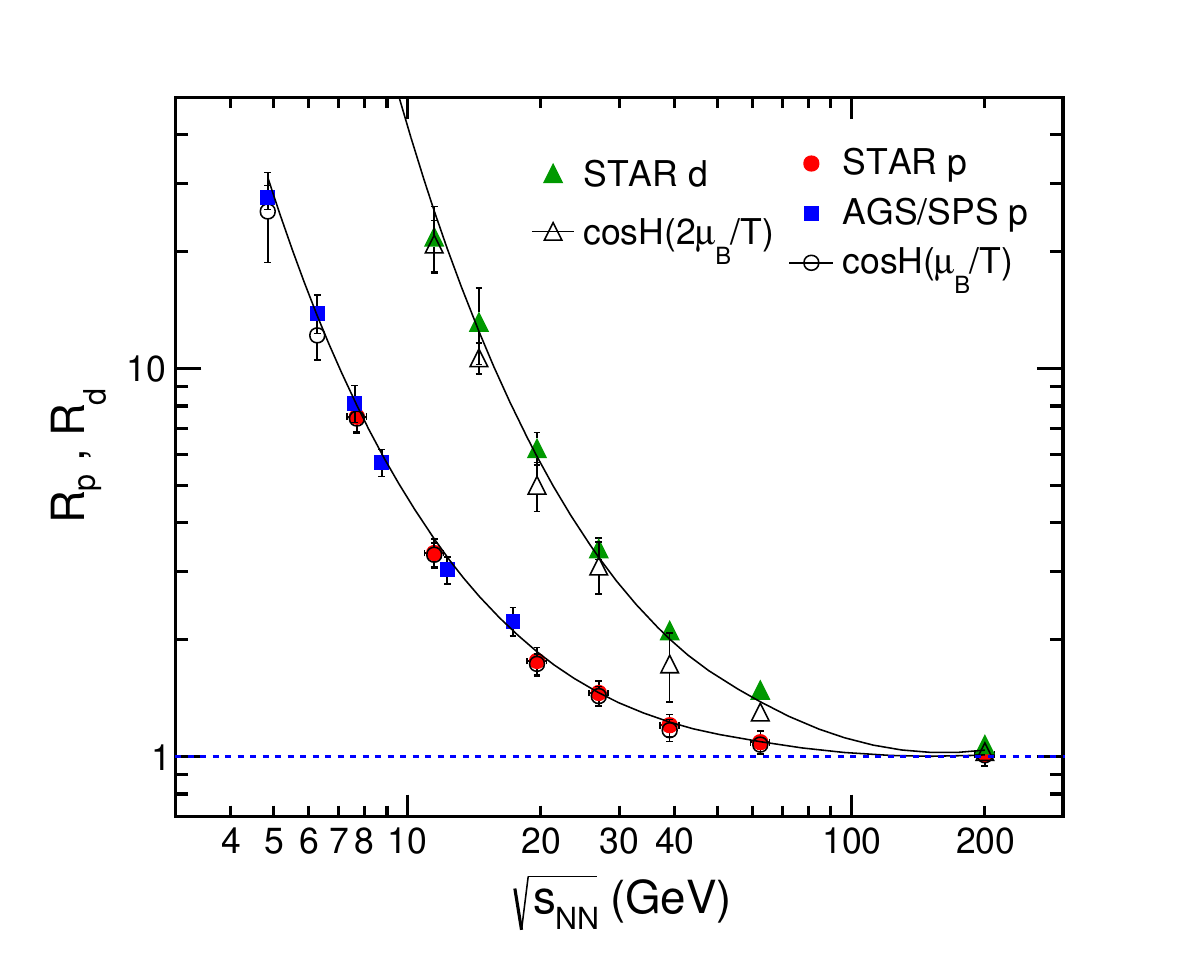}
\includegraphics[width=0.45\textwidth]{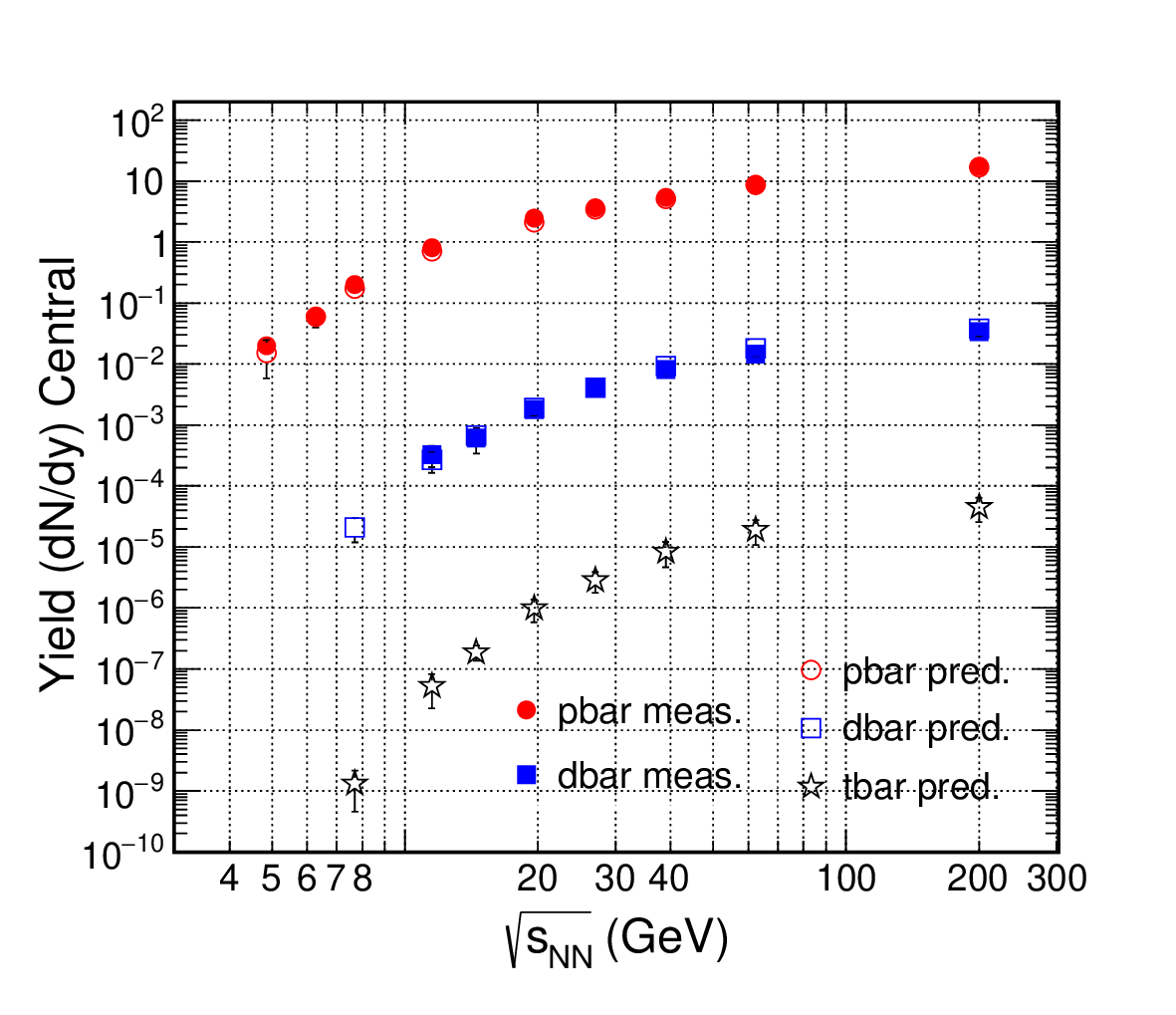}
  \caption{Top: Ratios $R_p$ and $R_d$, and their comparison with model
  predictions using $\mu_B/T$ as per Eqs.~(\ref{eq:P}) and (\ref{eq:D}). Bottom: Prediction of anti-nuclei 
  $\bar{p}$, $\bar{d}$, and $\bar{t}$ using measurements of nuclei and Eqs.\ 
  (\ref{eq:P}), (\ref{eq:D}), and (\ref{eq:T}) as explained in the text. 
  Unfilled symbols represent the predictions while the solid symbols represent
  the published measured yields. }
\label{fig:nuclei}
\end{figure}

For deuterons ($B_d=2$) and tritons ($B_t=3$), Eq.\ (\ref{eq:single_ratio}) can be written as
\begin{eqnarray}
R_d &\equiv& \frac{n_d + n_{\overline{d}}}{2\sqrt{n_d n_{\overline{d}}}} \approx \cosh\left(\frac{2\mu_B}{T}\right), \label{eq:D}\\
R_t &\equiv& \frac{n_t + n_{\overline{t}}}{2\sqrt{n_t n_{\bar{t}}}} \approx
      \cosh\left(\frac{3\mu_B}{T}\right). \label{eq:T}
\end{eqnarray}
We have neglected $\mu_Q/T$ because it is very small compared to $\mu_B/T$.

The upper panel of Fig.\ \ref{fig:nuclei} shows $R_p$ and $R_d$ as functions of $\sqrt{s_{NN}}$. The solid symbols represent values determined using measured 
experimental yields~\cite{NA49:2006gaj, Ahmad:1998sg, STAR:2017sal,STAR:2022hbp, STAR:2019sjh}.
The unfilled symbols represent predictions using the values of $\mu_B/T$ in Section \ref{sec:newfits}, while the smooth lines represent predictions using the values of $\mu_B/T$ from \cite{STAR:2017sal}.

It is observed that $R_p$ and $R_d$ approach unity with increasing energy, since $\mu/T$ decreases rapidly with increasing $\sqrt{s_{NN}}$ for reasons that we have discussed in Section \ref{sec:newfits}. For any given energy, the ratio deviates further from unity the more baryons it has, as shown in Eqs.\ (\ref{eq:D}) and (\ref{eq:T}).

Since Eqs.\ (\ref{eq:P}) and (\ref{eq:D}) seem to work well, one can invert these equations and use the measured yields of nuclei to give predictions of anti-nuclei yields where measurements are absent or statistically limited. We compare these model predictions with direct measurements of yields in the bottom plot of Figure \ref{fig:nuclei}. The predicted anti-proton and anti-deuteron yields agree well with the experimental measurements, as is to be expected from the panel above that. Note that $\bar d$ yields are not available at $\sqrt{s_{NN}}=7.7$ GeV, so the value given in the figure is a true prediction. We have also given a similar prediction for $\bar t$ using Eq.\ (\ref{eq:T}) and measurements of tritons~\cite{STAR:2022hbp}.
These are also genuine predictions of the model and can be tested in the future.

\section{Conclusions}\label{sec:conclusions}

Double ratios of particle and antiparticle yields have a very simple form in thermal models, as shown in Eq.\ (\ref{eq:lpxk}), when the particle mass is much larger than $T$ and $\mu_Q/T$ is small. This makes the equality of certain double ratios into simple tests of thermal models. We demonstrate such a test in Figure \ref{fig:RatioVsEnergy}. Clearly, quick tests such as this, or ones involving susceptibilities, as in Eq.\ (\ref{eq:cumulants}), would be important as heavy-ion collision energies are further lowered in the future. 

We build a different variable, $R_h$ defined in Eq.\ (\ref{eq:single_ratio}), which can be used to extract $\mu_{B,S,Q}/T$ with minimal computational work. We demonstrate the use of such an approach by showing the extraction of $\mu_Q/T$ using BES-I data from STAR. The data already show certain interesting trends with varying $N_{\rm part}$ and $\sqrt{s_{NN}}$. Certainly, using BES-II data will give better control on the errors in this determination purely through the increased statistics. Furthermore, at lower energy and higher event rates, one can envisage a more detailed study of $\mu_Q/t$. This could have implications on the extrapolation of heavy-ion data to neutron stars, especially when $\mu_Q$ is traded for the isospin chemical potential.

We have demonstrated that this method agrees with the results of the fits reported by STAR using the same input data. We also verified that the method correctly predicts $R_\Omega$. We also presented a parametrization for the change of $T$, $\mu_B$, and $\mu_S$ with $\sqrt{s_{NN}}$. For $\mu_S$, this is the first time that the energy dependence has been parametrized. These parameters allow us to predict the $T$, $\mu_B$, and $\mu_S$ at lower energies (AGS, SPS and below) and at energies where they are not yet estimated in BES (for example, at $\sqrt{s_{NN}}=14.5$ GeV).

The use of the variable $R_h$, defined in Eq.\ (\ref{eq:single_ratio}), also extends naturally to light (anti-)nuclei. This extension is agnostic to whether these bound states are created by statistical hadronization or a coalescence mechanism, as we argued in Section \ref{sec:nuclei}. We tested this idea using $R_d$ constructed from deuteron and anti-deuteron yields. Then we inverted this argument to predict the anti-deuteron yield at $\sqrt{s_{NN}}=7.7$ GeV, where it has not yet been reported. We also used this method to predict anti-tritium yields across a range of energies. This could be tested in the future.

In exploring the phase diagram of QCD, the focus has moved to regions with high $\mu_B$ and low $T$. This is the domain of experiments with low beam energy and high statistics. At the same time, $\mu_S$ and $\mu_Q$ are also needed with precision so that the knowledge gleaned from heavy-ion collisions can be extrapolated to neutron stars. We have demonstrated that building new combinations of experimental observables can explore more details than global fits have revealed in the past.

\section*{Acknowledgements}
The authors thank S. Chatterjee for interesting and useful discussions. L.K. acknowledges the financial support from Research Grant No. SR/MF/PS-02/2021-PU (E-37120) of the Department of Science and Technology, Government of India.
N.S. acknowledges the support from DU-IOE grant Ref. No./IoE/2025-26/12/FRP.



\bibliographystyle{elsarticle-num}
\bibliography{references_cprs}

\begin{thebibliography}{10}
\expandafter\ifx\csname url\endcsname\relax
  \def\url#1{\texttt{#1}}\fi
\expandafter\ifx\csname urlprefix\endcsname\relax\def\urlprefix{URL }\fi
\expandafter\ifx\csname href\endcsname\relax
  \def\href#1#2{#2} \def\path#1{#1}\fi

\bibitem{Hagedorn:1980kb}
R.~Hagedorn, J.~Rafelski, {Hot Hadronic Matter and Nuclear Collisions}, Phys.
  Lett. B 97 (1980) 136.
\newblock \href {https://doi.org/10.1016/0370-2693(80)90566-3}
  {\path{doi:10.1016/0370-2693(80)90566-3}}.

\bibitem{Hagedorn:1984uy}
R.~Hagedorn, K.~Redlich, {Statistical Thermodynamics in Relativistic Particle
  and Ion Physics: Canonical or Grand Canonical?}, Z. Phys. C 27 (1985) 541.
\newblock \href {https://doi.org/10.1007/BF01436508}
  {\path{doi:10.1007/BF01436508}}.

\bibitem{Becattini:2016xct}
F.~Becattini, J.~Steinheimer, R.~Stock, M.~Bleicher, {Hadronization conditions
  in relativistic nuclear collisions and the QCD pseudo-critical line}, Phys.
  Lett. B 764 (2017) 241--246.
\newblock \href {http://arxiv.org/abs/1605.09694} {\path{arXiv:1605.09694}},
  \href {https://doi.org/10.1016/j.physletb.2016.11.033}
  {\path{doi:10.1016/j.physletb.2016.11.033}}.

\bibitem{Bazavov:2017dus}
A.~Bazavov, et~al., {The QCD Equation of State to $\mathcal{O}(\mu_B^6)$ from
  Lattice QCD}, Phys. Rev. D 95~(5) (2017) 054504.
\newblock \href {http://arxiv.org/abs/1701.04325} {\path{arXiv:1701.04325}},
  \href {https://doi.org/10.1103/PhysRevD.95.054504}
  {\path{doi:10.1103/PhysRevD.95.054504}}.

\bibitem{Chatterjee:2013yga}
S.~Chatterjee, R.~M. Godbole, S.~Gupta, {Strange freezeout}, Phys. Lett. B 727
  (2013) 554--557.
\newblock \href {http://arxiv.org/abs/1306.2006} {\path{arXiv:1306.2006}},
  \href {https://doi.org/10.1016/j.physletb.2013.11.008}
  {\path{doi:10.1016/j.physletb.2013.11.008}}.

\bibitem{Becattini:2003wp}
F.~Becattini, M.~Gazdzicki, A.~Keranen, J.~Manninen, R.~Stock, {Chemical
  equilibrium in nucleus nucleus collisions at relativistic energies}, Phys.
  Rev. C 69 (2004) 024905.
\newblock \href {http://arxiv.org/abs/hep-ph/0310049}
  {\path{arXiv:hep-ph/0310049}}, \href
  {https://doi.org/10.1103/PhysRevC.69.024905}
  {\path{doi:10.1103/PhysRevC.69.024905}}.

\bibitem{Manninen:2008mg}
J.~Manninen, F.~Becattini, {Chemical freeze-out in ultra-relativistic heavy ion
  collisions at s(NN)**(1/2) = 130 and 200-GeV}, Phys. Rev. C 78 (2008) 054901.
\newblock \href {http://arxiv.org/abs/0806.4100} {\path{arXiv:0806.4100}},
  \href {https://doi.org/10.1103/PhysRevC.78.054901}
  {\path{doi:10.1103/PhysRevC.78.054901}}.

\bibitem{Becattini:2010sk}
F.~Becattini, P.~Castorina, A.~Milov, H.~Satz, {A Comparative analysis of
  statistical hadron production}, Eur. Phys. J. C 66 (2010) 377--386.
\newblock \href {http://arxiv.org/abs/0911.3026} {\path{arXiv:0911.3026}},
  \href {https://doi.org/10.1140/epjc/s10052-010-1265-y}
  {\path{doi:10.1140/epjc/s10052-010-1265-y}}.

\bibitem{Andronic_2018}
A.~Andronic, P.~Braun-Munzinger, K.~Redlich, J.~Stachel,
  \href{http://dx.doi.org/10.1038/s41586-018-0491-6}{Decoding the phase
  structure of qcd via particle production at high energy}, Nature 561~(7723)
  (2018) 321–330.
\newblock \href {https://doi.org/10.1038/s41586-018-0491-6}
  {\path{doi:10.1038/s41586-018-0491-6}}.
\newline\urlprefix\url{http://dx.doi.org/10.1038/s41586-018-0491-6}

\bibitem{Braun-Munzinger:2003pwq}
P.~Braun-Munzinger, K.~Redlich, J.~Stachel, {Particle production in heavy ion
  collisions} (2003) 491--599\href {http://arxiv.org/abs/nucl-th/0304013}
  {\path{arXiv:nucl-th/0304013}}, \href
  {https://doi.org/10.1142/9789812795533_0008}
  {\path{doi:10.1142/9789812795533_0008}}.

\bibitem{Andronic:2005yp}
A.~Andronic, P.~Braun-Munzinger, J.~Stachel, {Hadron production in central
  nucleus-nucleus collisions at chemical freeze-out}, Nucl. Phys. A 772 (2006)
  167--199.
\newblock \href {http://arxiv.org/abs/nucl-th/0511071}
  {\path{arXiv:nucl-th/0511071}}, \href
  {https://doi.org/10.1016/j.nuclphysa.2006.03.012}
  {\path{doi:10.1016/j.nuclphysa.2006.03.012}}.

\bibitem{Braun-Munzinger:2015hba}
P.~Braun-Munzinger, V.~Koch, T.~Sch{\"a}fer, J.~Stachel, {Properties of hot and
  dense matter from relativistic heavy ion collisions}, Phys. Rept. 621 (2016)
  76--126.
\newblock \href {http://arxiv.org/abs/1510.00442} {\path{arXiv:1510.00442}},
  \href {https://doi.org/10.1016/j.physrep.2015.12.003}
  {\path{doi:10.1016/j.physrep.2015.12.003}}.

\bibitem{STAR:2017sal}
L.~Adamczyk, et~al., {Bulk Properties of the Medium Produced in Relativistic
  Heavy-Ion Collisions from the Beam Energy Scan Program}, Phys. Rev. C 96~(4)
  (2017) 044904.
\newblock \href {http://arxiv.org/abs/1701.07065} {\path{arXiv:1701.07065}},
  \href {https://doi.org/10.1103/PhysRevC.96.044904}
  {\path{doi:10.1103/PhysRevC.96.044904}}.

\bibitem{Sharma:2018jqf}
N.~Sharma, J.~Cleymans, B.~Hippolyte, M.~Paradza, {A Comparison of p-p, p-Pb,
  Pb-Pb Collisions in the Thermal Model: Multiplicity Dependence of Thermal
  Parameters}, Phys. Rev. C 99~(4) (2019) 044914.
\newblock \href {http://arxiv.org/abs/1811.00399} {\path{arXiv:1811.00399}},
  \href {https://doi.org/10.1103/PhysRevC.99.044914}
  {\path{doi:10.1103/PhysRevC.99.044914}}.

\bibitem{Sharma:2018owb}
N.~Sharma, J.~Cleymans, L.~Kumar, {Thermal model description of
  p\textendash{}Pb collisions at $\sqrt{s_{NN}} = 5.02$ TeV}, Eur. Phys. J. C
  78~(4) (2018) 288.
\newblock \href {http://arxiv.org/abs/1802.07972} {\path{arXiv:1802.07972}},
  \href {https://doi.org/10.1140/epjc/s10052-018-5767-3}
  {\path{doi:10.1140/epjc/s10052-018-5767-3}}.

\bibitem{Braun-Munzinger:2024ybd}
P.~Braun-Munzinger, K.~Redlich, N.~Sharma, J.~Stachel, {Emergence of new
  systematics for open charm production in high energy collisions}, JHEP 04
  (2025) 058.
\newblock \href {http://arxiv.org/abs/2408.07496} {\path{arXiv:2408.07496}},
  \href {https://doi.org/10.1007/JHEP04(2025)058}
  {\path{doi:10.1007/JHEP04(2025)058}}.

\bibitem{Wheaton:2004qb}
S.~Wheaton, J.~Cleymans, {THERMUS: A Thermal model package for ROOT}, Comput.
  Phys. Commun. 180 (2009) 84--106.
\newblock \href {http://arxiv.org/abs/hep-ph/0407174}
  {\path{arXiv:hep-ph/0407174}}, \href
  {https://doi.org/10.1016/j.cpc.2008.08.001}
  {\path{doi:10.1016/j.cpc.2008.08.001}}.

\bibitem{Mark:2024hrv}
D.~K. Mark, F.~Surace, A.~Elben, A.~L. Shaw, J.~Choi, G.~Refael, M.~Endres,
  S.~Choi, {Maximum Entropy Principle in Deep Thermalization and in
  Hilbert-Space Ergodicity}, Phys. Rev. X 14~(4) (2024) 041051.
\newblock \href {http://arxiv.org/abs/2403.11970} {\path{arXiv:2403.11970}},
  \href {https://doi.org/10.1103/PhysRevX.14.041051}
  {\path{doi:10.1103/PhysRevX.14.041051}}.

\bibitem{Gupta:2022phu}
S.~Gupta, D.~Mallick, D.~K. Mishra, B.~Mohanty, N.~Xu, {Limits of
  thermalization in relativistic heavy ion collisions}, Phys. Lett. B 829
  (2022) 137021.
\newblock \href {https://doi.org/10.1016/j.physletb.2022.137021}
  {\path{doi:10.1016/j.physletb.2022.137021}}.

\bibitem{E-802:1998xum}
L.~Ahle, et~al., {Particle production at high baryon density in central Au+Au
  reactions at 11.6A GeV/c}, Phys. Rev. C 57~(2) (1998) R466--R470.
\newblock \href {https://doi.org/10.1103/PhysRevC.57.R466}
  {\path{doi:10.1103/PhysRevC.57.R466}}.

\bibitem{Ahmad:1991nv}
S.~Ahmad, et~al., {Lambda production by 11.6-A/GeV/c Au beam on Au target},
  Phys. Lett. B 382 (1996) 35--39, [Erratum: Phys.Lett.B 386, 496--496 (1996)].
\newblock \href {https://doi.org/10.1016/0370-2693(96)00642-9}
  {\path{doi:10.1016/0370-2693(96)00642-9}}.

\bibitem{Albergo:2002tn}
S.~Albergo, et~al., {Lambda spectra in 11.6-A-GeV/c Au Au collisions}, Phys.
  Rev. Lett. 88 (2002) 062301.
\newblock \href {https://doi.org/10.1103/PhysRevLett.88.062301}
  {\path{doi:10.1103/PhysRevLett.88.062301}}.

\bibitem{Ahmad:1998sg}
S.~Ahmad, B.~E. Bonner, S.~V. Efremov, G.~S. Mutchler, E.~D. Platner, H.~W.
  Themann, {Nuclear matter expansion parameters from the measurement of
  differential multiplicities for lambda production in central au+au collisions
  at ags}, Nucl. Phys. A 636 (1998) 507--524.
\newblock \href {http://arxiv.org/abs/nucl-ex/9803006}
  {\path{arXiv:nucl-ex/9803006}}, \href
  {https://doi.org/10.1016/S0375-9474(98)00218-8}
  {\path{doi:10.1016/S0375-9474(98)00218-8}}.

\bibitem{E917:2001eko}
B.~B. Back, et~al., {Anti-lambda production in Au+Au collisions at
  11.7-AGeV/c}, Phys. Rev. Lett. 87 (2001) 242301.
\newblock \href {http://arxiv.org/abs/nucl-ex/0101008}
  {\path{arXiv:nucl-ex/0101008}}, \href
  {https://doi.org/10.1103/PhysRevLett.87.242301}
  {\path{doi:10.1103/PhysRevLett.87.242301}}.

\bibitem{NA49:2006gaj}
C.~Alt, et~al., {Energy and centrality dependence of anti-p and p production
  and the anti-Lambda/anti-p ratio in Pb+Pb collisions between 20/A-GeV and
  158/A-Gev}, Phys. Rev. C 73 (2006) 044910.
\newblock \href {https://doi.org/10.1103/PhysRevC.73.044910}
  {\path{doi:10.1103/PhysRevC.73.044910}}.

\bibitem{NA49:2008ysv}
C.~Alt, et~al., {Energy dependence of Lambda and Xi production in central Pb+Pb
  collisions at A-20, A-30, A-40, A-80, and A-158 GeV measured at the CERN
  Super Proton Synchrotron}, Phys. Rev. C 78 (2008) 034918.
\newblock \href {http://arxiv.org/abs/0804.3770} {\path{arXiv:0804.3770}},
  \href {https://doi.org/10.1103/PhysRevC.78.034918}
  {\path{doi:10.1103/PhysRevC.78.034918}}.

\bibitem{Abelev:2008ab}
B.~Abelev, et~al., {Systematic Measurements of Identified Particle Spectra in
  $p p, d^+$ Au and Au+Au Collisions from STAR}, Phys. Rev. C 79 (2009) 034909.
\newblock \href {http://arxiv.org/abs/0808.2041} {\path{arXiv:0808.2041}},
  \href {https://doi.org/10.1103/PhysRevC.79.034909}
  {\path{doi:10.1103/PhysRevC.79.034909}}.

\bibitem{cleymans:2005xv}
J.~Cleymans, H.~Oeschler, K.~Redlich, S.~Wheaton, {Comparison of chemical
  freeze-out criteria in heavy-ion collisions}, Phys. Rev. C 73 (2006) 034905.
\newblock \href {http://arxiv.org/abs/hep-ph/0511094}
  {\path{arXiv:hep-ph/0511094}}, \href
  {https://doi.org/10.1103/PhysRevC.73.034905}
  {\path{doi:10.1103/PhysRevC.73.034905}}.

\bibitem{cleymans:2011pe}
J.~Cleymans, S.~Kabana, I.~Kraus, H.~Oeschler, K.~Redlich, N.~Sharma,
  {Antimatter production in proton-proton and heavy-ion collisions at
  ultrarelativistic energies}, Phys. Rev. C 84 (2011) 054916.
\newblock \href {http://arxiv.org/abs/1105.3719} {\path{arXiv:1105.3719}},
  \href {https://doi.org/10.1103/PhysRevC.84.054916}
  {\path{doi:10.1103/PhysRevC.84.054916}}.

\bibitem{Andronic:2009jd}
A.~Andronic, P.~Braun-Munzinger, J.~Stachel, {The Horn, the hadron mass
  spectrum and the QCD phase diagram: The Statistical model of hadron
  production in central nucleus-nucleus collisions}, Nucl. Phys. A 834 (2010)
  237C--240C.
\newblock \href {http://arxiv.org/abs/0911.4931} {\path{arXiv:0911.4931}},
  \href {https://doi.org/10.1016/j.nuclphysa.2009.12.048}
  {\path{doi:10.1016/j.nuclphysa.2009.12.048}}.

\bibitem{STAR:2025xxf}
{Identified charged hadron production in Au+Au collisions at
  $\sqrt{s_\mathrm{NN}}$ = 54.4 GeV with the STAR detector} (12 2025).
\newblock \href {http://arxiv.org/abs/2512.06415} {\path{arXiv:2512.06415}}.

\bibitem{Sharma:2022poi}
N.~Sharma, L.~Kumar, P.~M. Lo, K.~Redlich, {Light-nuclei production in pp and
  pA collisions in the baryon canonical ensemble approach}, Phys. Rev. C
  107~(5) (2023) 054903.
\newblock \href {http://arxiv.org/abs/2210.15617} {\path{arXiv:2210.15617}},
  \href {https://doi.org/10.1103/PhysRevC.107.054903}
  {\path{doi:10.1103/PhysRevC.107.054903}}.

\bibitem{STAR:2022hbp}
M.~Abdulhamid, et~al., {Beam Energy Dependence of Triton Production and Yield
  Ratio ($\mathrm{N}_t \times \mathrm{N}_p/\mathrm{N}_d^2$) in Au+Au Collisions
  at RHIC}, Phys. Rev. Lett. 130 (2023) 202301.
\newblock \href {http://arxiv.org/abs/2209.08058} {\path{arXiv:2209.08058}},
  \href {https://doi.org/10.1103/PhysRevLett.130.202301}
  {\path{doi:10.1103/PhysRevLett.130.202301}}.

\bibitem{STAR:2019sjh}
J.~Adam, et~al., {Beam energy dependence of (anti-)deuteron production in Au +
  Au collisions at the BNL Relativistic Heavy Ion Collider}, Phys. Rev. C
  99~(6) (2019) 064905.
\newblock \href {http://arxiv.org/abs/1903.11778} {\path{arXiv:1903.11778}},
  \href {https://doi.org/10.1103/PhysRevC.99.064905}
  {\path{doi:10.1103/PhysRevC.99.064905}}.

\end{thebibliography}

\end{document}